\documentclass[aps,prb,twocolumn,reprint]{revtex4-1}
\usepackage{amsmath,amssymb,bm,braket,bbm}
\usepackage[pdftex]{graphicx}
\usepackage{color}
\usepackage{comment}
\usepackage{braket}
\usepackage{mathtools}

\usepackage{ulem}
\definecolor{Green}{rgb}{0,0.5,0.15}

\usepackage{docmute}

\usepackage{mathptmx}
\usepackage{accents}
\newcommand\ddd[2]{\accentset{\circ}{#1}_{#2}}

\newcommand{\clam}{\Lambda}

\newcommand{\paren}[1]{\left(#1\right)}
\newcommand{\bck}[1]{\left[#1\right]}
\newcommand{\bce}[1]{\left\{#1\right\}}

\def\LL{{\mathrm{L}}}
\def\RR{{\mathrm{R}}}

\def\ssp{4pt}

\begin{document}

\title{
Generalized hydrodynamics study 
of the one-dimensional Hubbard model: \\
Stationary clogging and proportionality of spin, charge, and energy currents
}

\author{Yuji Nozawa and Hirokazu Tsunetsugu}
\affiliation{
The Institute for Solid State Physics, 
The University of Tokyo, Kashiwanoha 5-1-5, Chiba 277-8581, Japan}

\date{\today}

\begin{abstract}
In our previous work~[Nozawa and Tsunetsugu, 
Phys.~Rev.~B {\bf 101}, 035121 (2020)], we studied the quench dynamics 
in the one-dimensional Hubbard model 
based on the generalized hydrodynamics theory 
for a partitioning protocol 
and showed the presence of a \textit{clogging} phenomenon.   
Clogging is a phenomenon where vanishing charge current 
coexists with nonzero energy current, 
and we found it for the initial conditions 
that the left half of the system
is prepared to be half filling at high temperatures 
with the right half being empty.  
Clogging occurs at all the sites in the left half
and lasts for a time proportional to its distance 
from the connection point.  
In this paper, we use various different 
initial conditions and discuss two issues.   
The first issue is the possibility of clogging 
in a stationary state.  
When the electron density in the right half is initially set 
nonzero, we found that the left half-filled part expands 
for various sets of parameters in the initial condition.   
This means that the clogging phenomenon 
occurs at all the sites in the long-time stationary state, 
and we also discuss its origin. 
In addition, stationary clogging is accompanied by a \textit{back current}, namely, particle density current flows towards the high-density region. 
We also found that spin clogging occurs for some initial conditions, 
i.e., the vanishing spin current coexists with nonzero energy current. 
The second issue is the proportionality of spin and charge currents.
We found two spatio-temporal regions 
where the current ratio is fixed to a nonzero constant.  
We numerically studied how the current ratio 
depends on various initial conditions. 
We also studied the ratio of charge and energy currents.  
\end{abstract}
\maketitle

\section{Introduction}
\label{sec:1} 
Understanding nonequilibrium phenomena in strongly
correlated systems is an important and challenging issue, and 
one-dimensional (1D) integrable models have attracted attention 
because their infinite number of conserved quantities play an important
role~\cite{Calabrese_2016}.  
Recently, the generalized hydrodynamics (GHD) theory was proposed by the authors of 
Refs.~\onlinecite{PhysRevX.6.041065,PhysRevLett.117.207201} 
for studying nonequilibrium dynamics of integrable models, 
and its experimental confirmation was 
demonstrated for a 1D Bose gas system~\cite{PhysRevLett.122.090601}. 
An infinite number of conserved quantities is also
important in GHD, as time-evolution equations are
formulated based on their continuity equations.
The GHD can describe the time evolution of spatially 
inhomogeneous systems, and partitioning is a frequently 
used protocol~\cite{rubin1971abnormal,
spohn1977stationary,%
bernard2012energy,Bernard2015,bhaseen2015energy,%
PhysRevX.6.041065,PhysRevLett.117.207201}, 
because the equations for the time evolution are simple in that case.  
Two semi-infinite parts in different thermal equilibria 
are connected at the origin $x=0$ and time $t=0$,
and the time evolution of the connected system is analyzed.  
By using this protocol, many aspects of nonequilibrium phenomena in
integrable models have been studied, e.g., the time dependence of
currents~\cite{PhysRevX.6.041065,PhysRevLett.117.207201,%
fagotti2016charges,PhysRevB.96.020403,doyon2017dynamics,%
PhysRevB.97.045407,PhysRevLett.120.045301,PhysRevB.97.081111,%
Bertini_2018,PhysRevLett.120.176801,10.21468/SciPostPhys.4.6.045,%
PhysRevB.98.075421,PhysRevB.99.014305,PhysRevB.99.174203,%
doi:10.1063/1.5096892,Bulchandani_2019},
Drude weights~\cite{PhysRevLett.119.020602,PhysRevB.96.081118,%
SciPostPhys.3.6.039},
entanglements~\cite{PhysRevB.97.245135,Bertini_ent,PhysRevB.99.045150,%
10.21468/SciPostPhys.7.1.005},
correlation functions of densities and currents~\cite{PhysRevB.96.115124,%
10.21468/SciPostPhys.5.5.054},
diffusive dynamics and diffusion constants~\cite{PhysRevLett.121.230602,PhysRevLett.121.160603,PhysRevB.98.220303,%
10.21468/SciPostPhys.6.4.049,PhysRevLett.122.127202,Gopalakrishnan16250,PhysRevB.102.115121}.

\begin{figure}[tb]
\centering \includegraphics[width=0.9\columnwidth,clip]{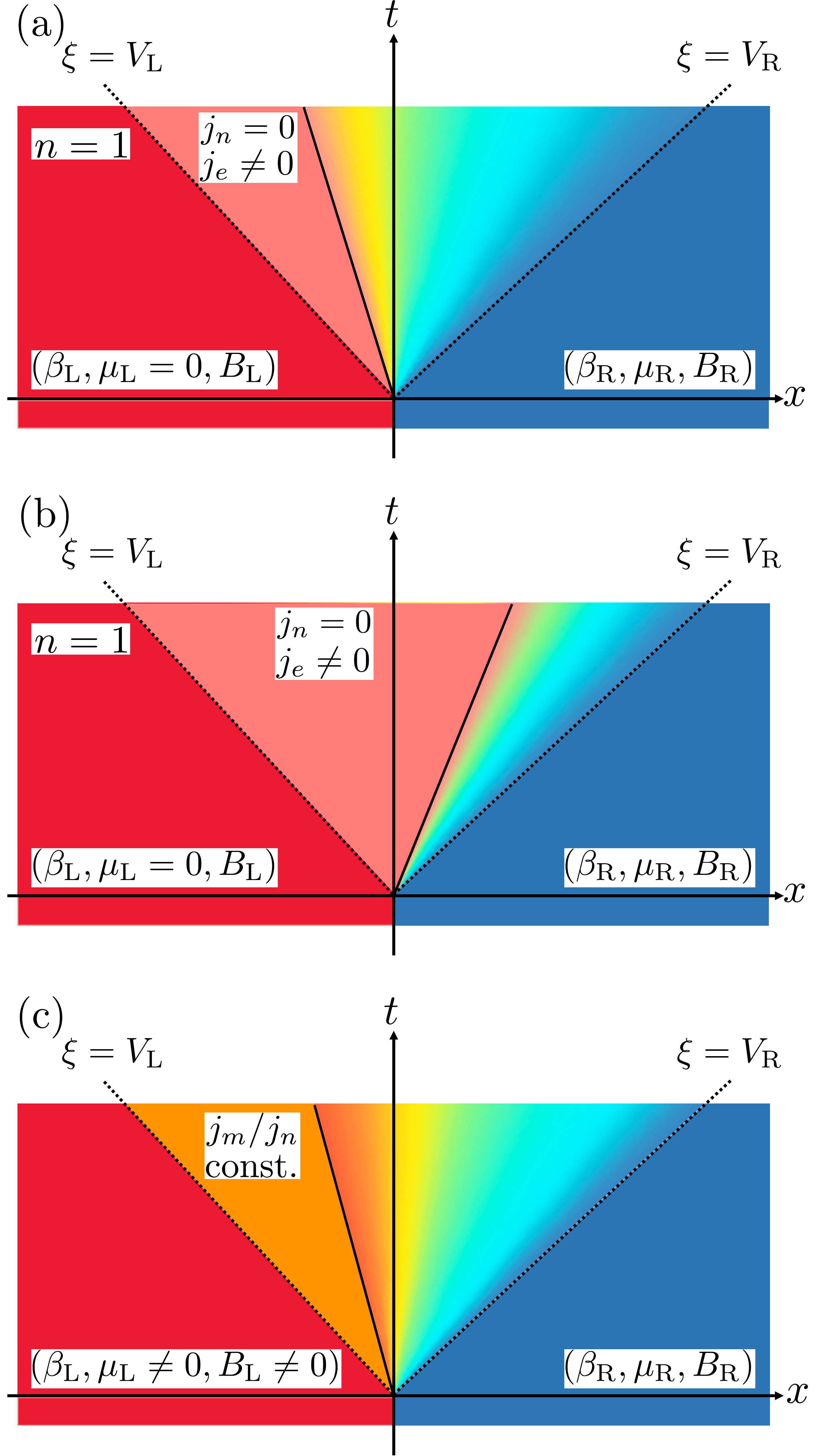}
\caption{
Schematic picture of partitioning protocol and 
nonequilibrium phenomena studied in this paper. 
(a) Charge clogged region, where $j_{n}=0$ and $j_{e}\neq 0$.
(b) Stationary charge clogging. 
The region of $n =1$ includes $\xi = 0$. 
(c) Proportionality of spin and charge currents when $\mu_{\LL}\neq 0$ and $B_{\LL}\neq 0$.
The ratio $j_{m}/j_{n}$ is fixed to a constant 
value in a region to the right of $\xi=V_{\LL}$. 
Outside the light cones, $\xi \leq V_{\LL}$ or $\xi \geq V_{\RR}$, 
local states are unchanged from the initial left or right thermal equilibrium states, respectively.}  \label{fig:spinghdsetup}
\end{figure}

The 1D Hubbard model is a canonical lattice model of strongly
correlated electrons and is exactly solvable through the nested Bethe
ansatz~\cite{PhysRevLett.19.1312, GAUDIN196755,
PhysRevLett.21.192.2,essler2005one}. 
Studying its nonequilibrium dynamics is very important to understand 
transport experiments in many quasi-1D systems including inorganic~\cite{PhysRevB.81.020405} and organic~\cite{PhysRevB.58.1261} compounds, 
quantum wires~\cite{Nature.397.598}, and fermionic cold atom systems~\cite{Boll1257}.
Ilievski and De Nardis formulated its GHD theory 
and also confirmed it by numerical calculations~\cite{PhysRevB.96.081118}.
In our previous work, we used their formulation with the partitioning
protocol and mainly studied charge and energy
currents~\cite{PhysRevB.101.035121}.
We found the existence of a region that has zero charge (spin) current
while nonzero energy current flows and named it
\textit{charge (spin) clogged region}
[see Fig.~\ref{fig:spinghdsetup}(a)]. 
We proved its existence for the cases 
that the left side of the initial state is at infinite temperature 
$\beta_{\LL} = 0$. 
We also numerically studied charge and energy currents 
in the cases of $\beta_{\LL} >0$ where 
the initial right state has no electron. 
In these calculations, clogging occurs at sites in the left half 
for a finite period of time that is proportional 
to the site position measured from the origin. 
It is an interesting question whether one can realize 
such a peculiar phenomenon as \textit{charge or spin clogging} 
in the stationary state, 
and if the answer is positive it is important to find its conditions 
as a theoretical prediction for experimental observations. 
A related important issue is the ratio of different kinds of currents, 
e.g., charge $( j_n )$, spin $( j_m )$, and energy $( j_e )$ currents, 
since it is an observable evidence of multiple types of quasiparticles.  
Our previous paper~\cite{PhysRevB.101.035121} mainly analyzed 
the ratio $j_e / j_n$, 
which is related to the Wiedemann-Franz law in thermal equilibrium~\cite{WF}, 
but $j_m / j_n$ was calculated only in the high-temperature limit.  
From the viewpoint of condensed matter physics, 
it is also important to see how the two currents $j_m$ and $j_n$ are 
related in the strongly correlated electron systems when a magnetic field is applied.

In this paper, to clarify these points, we will use the partitioning protocol with a wider range
of initial conditions and study the profiles of spin, 
charge, and energy currents.
We will mainly discuss two issues. 
The first issue is the possibility of expanding the charge clogged region
[see Fig.~\ref{fig:spinghdsetup}(b)].
We found that 
the left half-filled region expands, and charge clogging
occurs in the stationary state, 
when the initial right temperature is lower than the initial 
left one ($\beta_{\LL}<\beta_{\RR}$) or 
a magnetic field is applied in the initial right part ($B_{\RR}>0$). 
This will be discussed in Sec.~\ref{sec:4}.
The second issue is the proportionality of spin and charge currents~[see Fig.~\ref{fig:spinghdsetup}(c)], 
which was studied only in the high-temperature limit in our previous 
work~\cite{PhysRevB.101.035121}.
We will study this issue in Sec.~\ref{sec:5}
for the cases of finite temperature 
and show that there emerge two regions
where the ratio of spin current to charge current
is fixed to a constant value.

This paper is organized as follows. In Sec.~\ref{sec:2}, we introduce
the 1D Hubbard model and the GHD approach to it. In particular, we
describe how to calculate the profiles of densities and currents for the
partitioning protocol.  
In Sec.~\ref{sec:4}, we present the main results on the expansion of a
half-filled region.  We show initial conditions where stationary
charge clogging occurs and analyze the initial conditions
dependence of the existence of it.  
We also examine stationary spin clogging.  
In Sec.~\ref{sec:5}, we present the main results on 
the proportionality of spin and charge currents at finite
temperatures. To study the proportionality, we analyze the profiles of
the ratio of spin current to particle density current and their initial
conditions dependence.  We also analyze the profiles of the ratio of
energy current to particle density current.  
Finally, the conclusions
are given in Sec.~\ref{sec:6}.
 
\section{Model and Method}
\label{sec:2} 
Let us briefly summarize in this section the GHD approach to the 1D Hubbard
model~\cite{PhysRevB.96.081118,PhysRevB.101.035121}.
Throughout this paper, we will use the notations defined in
our previous work (Ref.~\onlinecite{PhysRevB.101.035121}).
Refer to that paper for more details of the calculations.

The Hamiltonian of the 1D Hubbard model on $L$ sites reads as 
\begin{align}
\hat{H}
=&-\sum_{j=1}^{L}\sum_{\sigma} 
\left[
\bigl( \hat{c}_{j,\sigma}^{\dagger}\hat{c}_{j+1,\sigma}+\mathrm{H.c.} \bigr)
+ (\mu + s_{\sigma} B ) \hat{n}_{j, \sigma}
\right]
\nonumber\\
&+4u \sum_{j=1}^{L}
\bck{ 
\bigl( \hat{n}_{j,\uparrow}-{\textstyle \frac{1}{2}} \bigr)
\bigl( \hat{n}_{j,\downarrow}-{\textstyle \frac{1}{2}} \bigr)-{\textstyle \frac{1}{4}}},
\label{eq:Hamiltonian}
\end{align}
where $\hat{c}_{j,\sigma}^{\dagger}$ and $\hat{c}_{j,\sigma}$ 
are the electron creation and annihilation operator, respectively, 
at site $j$ with spin 
$\sigma\  \in \{\uparrow,\downarrow \}$. 
$\hat{n}_{j,\sigma} \equiv 
\hat{c}^{\dagger}_{j,\sigma} \hat{c}_{j,\sigma}$ and 
$s_{\sigma}$ is defined as $s_{\uparrow}=1$ and $s_{\downarrow}=-1$.
We set the electron hopping amplitude to be unity, 
and use it as the unit of energy throughout this paper.
$\mu$ and $B$ are chemical potential and magnetic field, respectively.  
The Coulomb repulsion is parameterized by $u~ >0$, 
and the constant $-1/4$ in this term is included so as to
make the energy of the vacuum state zero. 

The partitioning protocol is shown in Fig.~\ref{fig:spinghdsetup}.
Initially, the system is divided into left and right parts,
and they are independently thermalized with different sets 
of parameters ($\beta_{\LL}$, $\mu_{\LL}$, $B_{\LL}$) and
($\beta_{\RR}$, $\mu_{\RR}$, $B_{\RR}$).  
At time $t=0$, the two part are connected at the origin $x=0$,
and we study the time evolution of the total system.
The initial particle density $n^{\LL(\RR)}$, magnetization
$m^{\LL(\RR)}$, and energy density $e^{\LL(\RR)}$
are controlled by the corresponding set of the parameters.  
Hereafter we consider the case of 
$\mu_{\mathrm{s}}\leq 0$ and $B_{\mathrm{s}}\geq 0$ for $\mathrm{s}=\LL,
\RR$, which means $n^{\LL(\RR)}\leq 1$ and $m^{\LL(\RR)}\geq 0$. 

The GHD theory describes a state by the distribution
functions of quasiparticles $\bce{\rho_{a}(w; x, t)}$,
and the time evolution is defined by their continuity equations.
Here, the integer label $a$ denotes the type of quasiparticles.
The first type corresponding to $a=0$ is called \textit{real} $k$.
They are scattering states of polarized electrons, and each state 
carries the electron charge $e(<0)$ and spin projection $1/2$.
The variable $w$ takes a real value $k$ and represents
charge momentum $(-\pi<k\leq\pi)$.
The second type corresponding to $a>0$ is called $\clam$-\textit{string}. 
They are either scattering states of spins $(a=1)$ or
bound states of spins $(a>1)$.
Each state carries spin projection $-a$.  
In this case, the variable $w=\clam$ represents 
the real part of complex spin rapidity $(-\infty<\clam<\infty)$.
The third type corresponding to $a<0$ is called  $k$-$\clam$ \textit{string}.
They are bound states of charges,
and each state carries charge $2|a|e$.
The variable $w=\clam$ now
represents the real part of complex charge rapidity. 

The distribution functions evolve in time following the continuity
equations~\cite{PhysRevX.6.041065,PhysRevLett.117.207201,PhysRevLett.125.070602} 
$
  \frac{\partial}{\partial t} \, \rho_{a}\paren{w;x,t} 
+ \frac{\partial}{\partial x} \, 
  \left[ \ddd{v}{a} (w;x,t) \rho_{a} (w;x,t) \right] 
  =0
$.   
Here, $\{ \ddd{v}{a} \}$ are the dressed
velocities~\cite{PhysRevLett.113.187203}, and
the reader should refer to 
Refs.~\onlinecite{PhysRevB.96.081118, PhysRevB.101.035121}
to know how to obtain them. 
Upon using the partitioning protocol, it is known that the solution of the continuity equations only
depends on the ray $\xi\equiv x/t$, and it is convenient to introduce
the filling functions $\bce{\vartheta_{a}(w; \xi)}$ to represent the
solution~\cite{PhysRevX.6.041065,PhysRevLett.117.207201}.
Once $\bce{\vartheta_{a}(w; \xi)}$ are obtained, the distribution
functions $\bce{\rho_{a}(w; \xi)}$ are calculated 
by solving the integral equations called the Takahashi
equations~\cite{10.1143/PTP.47.69}.
The solution of the filling functions are written as
\begin{align}
 \vartheta_{a} (w, \xi ) 
 =
 ~ 
   \Theta \left( \ddd{v}{a} ( w ,  \xi ) - \xi \right) 
   \vartheta_{a}^{\LL} (w)
   + \Theta \left( \xi - \ddd{v}{a} ( w ,  \xi ) \right) 
          \vartheta_{a}^{\RR} (w)
\label{eq:vartheta}
\end{align}
with Heaviside's step function $\Theta (x)$.  
$\{\vartheta_{a}^{\LL(\RR)} (w)\}$ are the
initial left (right) filling functions and obtained by solving
the integral equations for thermal equilibrium specified by
the set of parameters 
$(\beta_{\LL(\RR)},\mu_{\LL(\RR)},B_{\LL(\RR)})$, which are called the
thermodynamic Bethe ansatz (TBA) equations~\cite{10.1143/PTP.47.69}.
We note that 
the dressed velocities 
depend on the filling functions
and therefore both of them have to be determined self-consistently.
The above solution shows that the value of $\vartheta_{a} (w, \xi )$
is identical to its initial value either in the left or right part. 

Once the distribution functions $\bce{\rho_{a}(w;\xi)}$ and
the dressed velocities $\{\ddd{v}{a} (w ,\xi)\}$ are obtained,
they suffice to calculate densities and currents.
The particle density $n$, magnetization $m$, and
energy density $e$ and
their currents $j_{n}$, $j_{m}$, $j_{e}$ are given
by~\cite{PhysRevB.96.081118}
\begin{equation}
\left[
\begin{array}{c}
\displaystyle
n_{r} (\xi )
\\[\ssp ]  
\displaystyle
j_{r} (\xi ) 
\end{array}
\right]
=  \sum_a \int dw \, 
\left[
\begin{array}{c}
\displaystyle
1  \\[\ssp ]  
\displaystyle
\ddd{v}{a} (w , \xi ) 
\end{array}
\right]
f_{r,a} (w) 
\rho_a (w ,\xi ) , 
\label{eq3:density_component}
\end{equation}
where the label ($r=n, m, e$) distinguishes densities 
$n_n=n$, $n_m = m$, and $n_e = e$ and
corresponding currents.  
The weights are defined as 
\begin{align}
f_{n,a} (w)  &= \delta_{a,0}  + |a| - a, 
\\  
  f_{m,a} (w)  &=
  {\textstyle \frac12} \left( \delta_{a,0}  -  |a| - a \right) , 
\\ 
  f_{e,a} (w)  &=
 e_a (w).
\end{align}
Here, $e_{a}$ is the bare energy of the type-$a$ quasiparticle:  
\begin{align}
 e_{0} (k)&= -2 \cos k  - 2u , \nonumber\\
 e_{a < 0} (\clam ) &= 
4\operatorname{Re} \sqrt{1-(\clam + i a u)^2}
+4au, 
\label{eq:be}
\end{align}
and $e _{a>0} (\clam ) = 0$.
The symbol $\operatorname{Re}$ denotes the real part.

We define light cones $\xi_a^{\pm}$ for each string $a$ for later use.  
From Eq.~\eqref{eq:vartheta}, $\xi_a^{\pm}$ are defined as 
the minimum and maximum $\xi$-values on the intersection line 
of the two surfaces 
$z_1 (w, \xi )= \ddd{v}{a} ( w ,  \xi )$ and 
$z_2 (w, \xi ) = \xi$.  
The filling function $\vartheta_{a}(w,\xi)$ continuously 
varies inside the light cone $\xi_{a}^{-} < \xi < \xi_{a}^{+}$, 
while outside the light cone it is fixed to either 
$\vartheta_{a}^{\LL}(w)$ or $\vartheta_{a}^{\RR}(w)$.  
In addition, we also define
\begin{align}
V_{\LL}&\equiv\min_{a}\xi_{a}^{-} = \xi_{0}^{-} , & 
V_{\RR}&\equiv\max_{a}\xi_{a}^{+} = \xi_{0}^{+},
\nonumber\\
V_{\LL,1}&\equiv\min_{a\neq 0}\xi_{a}^{-}, & 
V_{\RR,1}&\equiv\max_{a\neq 0}\xi_{a}^{+},
\label{eq:defboundary}
\end{align}
and the first two are determined by real $k$ quasiparticles.  
The definition means that all the filling functions 
are equal to the initial equilibrium values in the left part
at $\xi\leq V_{\LL}$,
while those in the right part at $\xi \geq V_{\RR}$.
We note that $V_{\LL}$ does not depend on $(\beta_{\RR},\mu_{\RR},B_{\RR})$,
and \textit{vice versa}.
In the regions $V_{\LL} \leq \xi\leq V_{\LL ,1}$ and 
$V_{\RR ,1} \leq \xi\leq V_{\RR}$, 
only real-$k$ quasiparticles have a filling function 
different from the initial equilibrium values.  

Throughout this paper we set the repulsion $u=2$.  Approximations used for the
numerical calculations at finite temperatures are the same as in our
previous work~\cite{PhysRevB.101.035121}, where the cut-off concerning
the number of integral equations $a_{c}$ is
used~\cite{PhysRevB.65.165104}. The values of the densities and currents
shown in figures are extrapolated ones obtained from the calculations
for $a_{c}=36, 42, \text{and}~48$.

\section{Stationary clogging}
\label{sec:4} 

When the initial left state is set half-filling ($\mu_{\LL}=0$) 
at infinite or high temperatures, there emerges a 
\textit{charge clogged} region near the left end of 
the intermediate transient region
\begin{align}
    n (\xi)=1 , \ j_n (\xi)=0 , \ j_{e} (\xi) \not\equiv 0, 
   \quad 
   \bigl( V_{\LL} < {}^\forall \xi < \xi_{-\infty}^{-} \bigr).  
\end{align}
In our previous work~\cite{PhysRevB.101.035121}, 
the initial right state was an electron vacuum,  
and then $\xi_{-\infty}^{-} < 0$ for all the parameters examined. 
Therefore, the clogging phenomenon appears only at sites in the left half, 
and also it continues only for a limited time, which is proportional to 
the distance between the site position and the origin.  

\begin{figure}[tb]
\centering
\includegraphics[width=\columnwidth,clip]{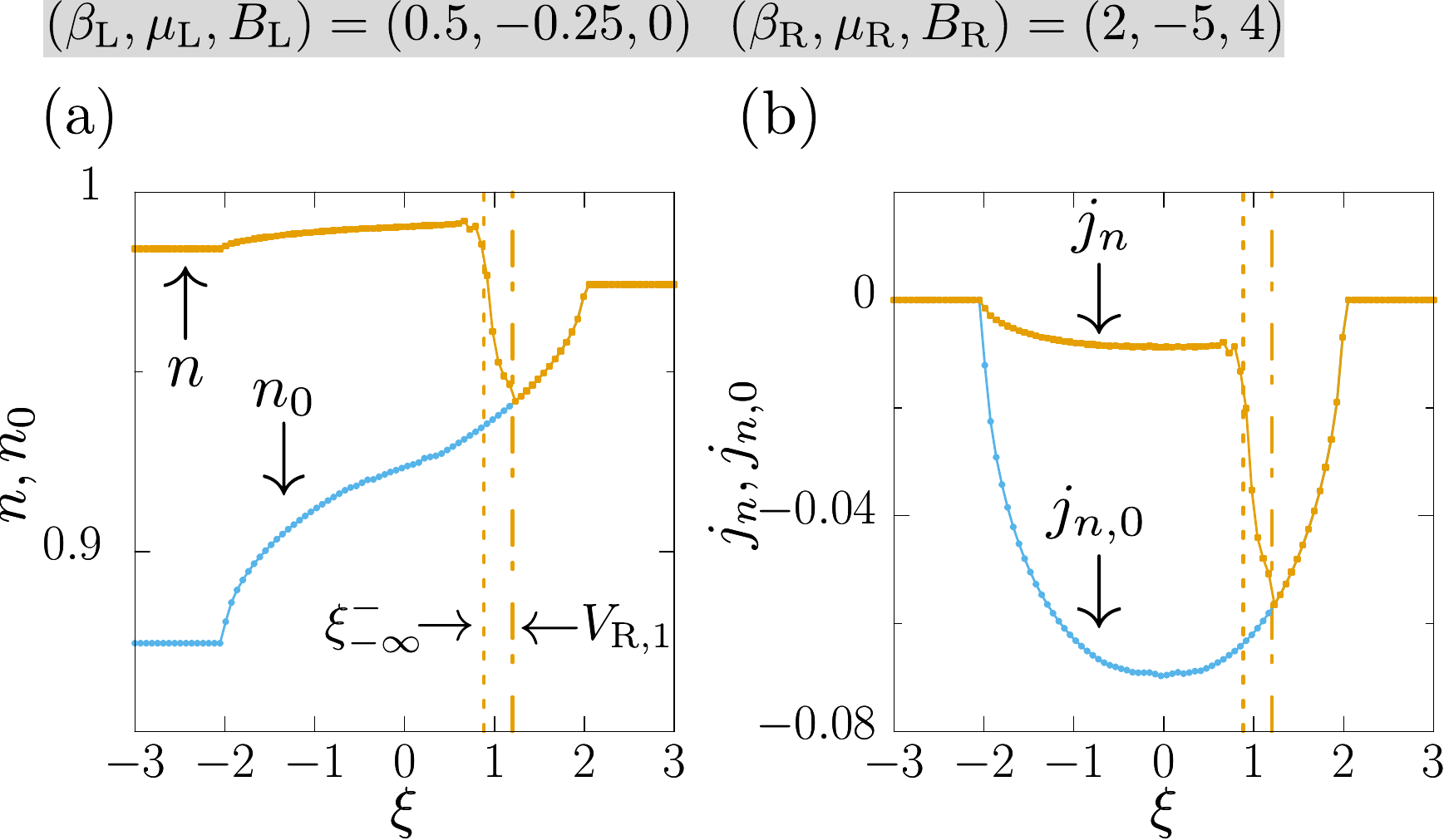}
\caption{ 
Profiles of (a) particle density $n(\xi )$ and 
(b) its current $j_n (\xi )$ when 
back current flows without charge clogging. 
}
\label{fig:backcurrent}
\end{figure}

We now examine if one can realize clogging in a stationary state 
by tuning initial conditions.   
Stationary values of physical quantities are those at $\xi =0$, 
and therefore the question is how to tune parameters for 
achieving $\xi_{-\infty}^{-} > 0$.  
We will show that a key point is the density of real-$k$ quasiparticles, 
$n_0 (\xi )$.  
In all the cases in this section, 
we will control the initial conditions in the right part, 
while the initial left state is prepared with $\beta_{\LL}=0.5$ 
and fixed at half-filling $\mu_{\LL} =0$ except for the data in Fig.~\ref{fig:backcurrent}.  

\begin{figure}[tb]
\centering
\includegraphics[width=\columnwidth,clip]{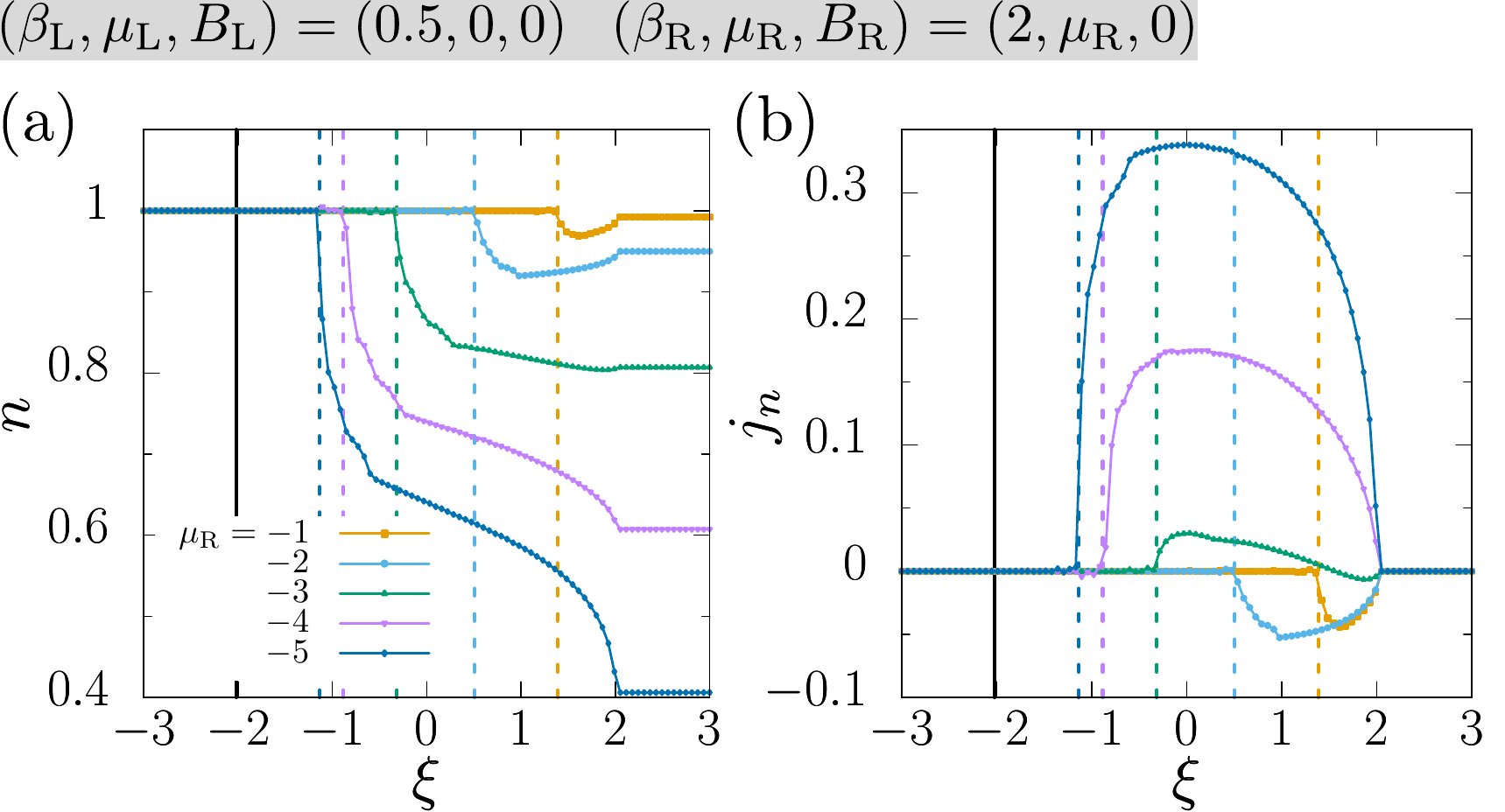}
\caption{
Profiles of (a) particle density $n(\xi )$ and (b) its current $j_n (\xi )$ 
for various values of $\mu_{\RR}$. 
Vertical dashed and solid lines are $\xi^{-}_{-\infty}$ and $V_{\LL}$, 
which are the borders of charge clogged region.}
\label{fig:njnclogged}
\end{figure}

Let us first set the initial temperature in the right part 
lower than the left part $\beta_{\RR}=2$, 
and control chemical potential in the range of 
$ -5 \le \mu_{\RR} \le -1$.  
Magnetic field is set zero in both parts $B_{\LL}=B_{\RR}=0$. 
Figure~\ref{fig:njnclogged} shows the profiles of
particle density $n(\xi)$ and its current $j_{n}(\xi)$. 
One should recall that 
$V_{\LL} \approx -2.0$ does not depend on $\mu_{\RR}$.  
The figure shows a clogging phenomenon for all $\mu_{\RR}$'s 
and the particle density is fixed to 1 in the clogging region. 
Its right end $\xi^{-}_{-\infty}$ moves to the right 
with increasing $\mu_{\RR}$, and the clogging region 
includes $\xi=0$ for the two largest values 
$\mu_{\LL}=-2$ and $-1$.
Thus, charge clogging occurs in the stationary state in these cases.  

Stationary clogging is accompanied by another interesting phenomenon, 
and that is \textit{back current}. 
Figure~\ref{fig:njnclogged}(b) shows a region 
where $j_n < 0$ for $\mu_{\LL}=-2$ and $-1$,  
namely particle density current flows towards the high-density region.  
This is related to a nonmonotonic behavior of $n(\xi)$ 
in Fig.~\ref{fig:njnclogged}(a).  
One can explain the presence of back current based on 
the continuity equation of particle density 
$\xi \partial_{\xi}n(\xi)=\partial_{\xi}j_{n}(\xi)$. 
Integrating this over the region $V_{\LL} \le \xi \le V_{\RR}$ 
with the boundary values 
$j_{n}(V_{\LL})=j_{n}(V_{\RR})=0$, one obtains 
\begin{align}
0 = \int_{V_{\LL}}^{V_{\RR}}d \xi \partial_{\xi} j_n (\xi) 
  = \int_{V_{\LL}}^{V_{\RR}}d \xi \xi \partial_{\xi}n(\xi) . 
\label{eq:ncon}
\end{align}
If the clogged region extends beyond $\xi = 0$, then 
$n (\xi) = 1 $ for ${}^\forall \xi \le 0$, 
and the above integral is rewritten as 
\begin{align}
0 = \int_{0}^{V_{\RR}}d \xi \xi \partial_{\xi}n(\xi)
  = - \int_{0}^{V_{\RR}}d \xi \delta n (\xi) , 
\label{eq:ncon2}
\end{align}
where $\delta n(\xi ) \equiv n(\xi) - n^{\RR}$ is density deviation.  
Since $ \delta n(\xi \approx 0) > 0 $,  
this integral means that there exists a finite-width region 
where $\delta n (\xi) <  0$.
At the right end,  $\delta n (V^{\RR}) =0$, and thus 
$n (\xi)$ should be nonmonotonic.  
Let us also examine particle current density.  
Just above $\xi^{-}_{-\infty} >0$, $\partial_\xi n$ is negative, and 
this leads to 
\begin{equation}
 j_n (\xi ) = \int_{\xi^{-}_{-\infty}}^{\xi} d\xi ' \, 
\xi ' \, \partial_{\xi '} n (\xi ' ) < 0 , 
\end{equation}
at least if $ 0 < \xi - \xi^{-}_{-\infty}  \ll 1$.  
Therefore, although the left part initially has a higher 
density of electrons, the particle current flows to the left 
in this region.  
This contrasts with the ordinary current flow driven by particle diffusion 
and may be called \textit{back current} in this sense.

\begin{figure}[tb]
\centering
\includegraphics[width=\columnwidth,clip]{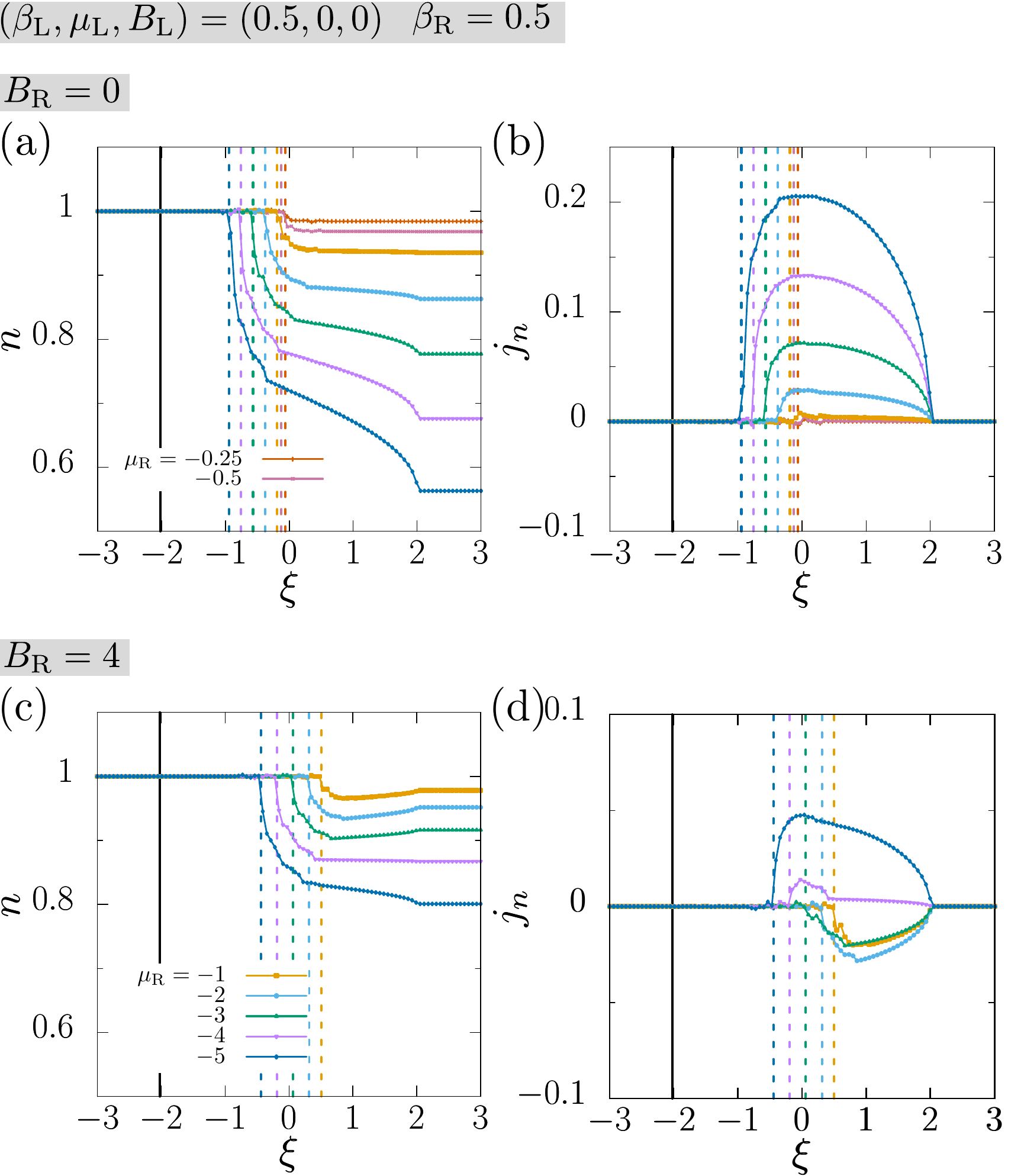}
\caption{
Effect of magnetic field in the right initial state $B_{\RR}$ 
on the profiles of $n(\xi)$ and $j_{n}(\xi)$. 
(a), (b) $B_{\RR} =0$ and $-5 \le \mu_{\RR} \le -0.25$, 
while (c), (d) $B_{\RR} =4$ and $-5 \le \mu_{\RR} \le -1$. }  
\label{fig:njncloggedmagnetic}
\end{figure}

We next consider the case of controlling $\mu_{\RR}$ when the initial temperature is identical in both 
parts $\beta_{\LL}=\beta_{\RR}=0.5$.  
As in the previous case, $\mu_{\LL}=0$ and $B_{\LL}=0$.
Figures~\ref{fig:njncloggedmagnetic}(a) and \ref{fig:njncloggedmagnetic}(b) 
show $n(\xi)$ and $j_{n}(\xi)$ at $B_{\RR}=0$ for 
$-5 \le \mu_{\RR} \le -0.25$.  
Since $\xi^{-}_{-\infty} < 0$ for all $\mu_{\RR}$'s, 
stationary charge clogging does not occur, 
and we try another type of control, i.e., applying magnetic field. 
Figures~\ref{fig:njncloggedmagnetic}(c) and \ref{fig:njncloggedmagnetic}(d) show 
$n(\xi)$ and $j_{n}(\xi)$ at $B_{\RR}=4$ 
for $-5 \le \mu_{\RR} \le -1$. 
The other initial conditions are the same as those in Figs.~\ref{fig:njncloggedmagnetic}(a) and \ref{fig:njncloggedmagnetic}(b).  
The result is that stationary charge clogging occurs 
for $\mu_{\RR}\geq -3$.

\begin{figure}[tb]
\centering
\includegraphics[width=\columnwidth,clip]{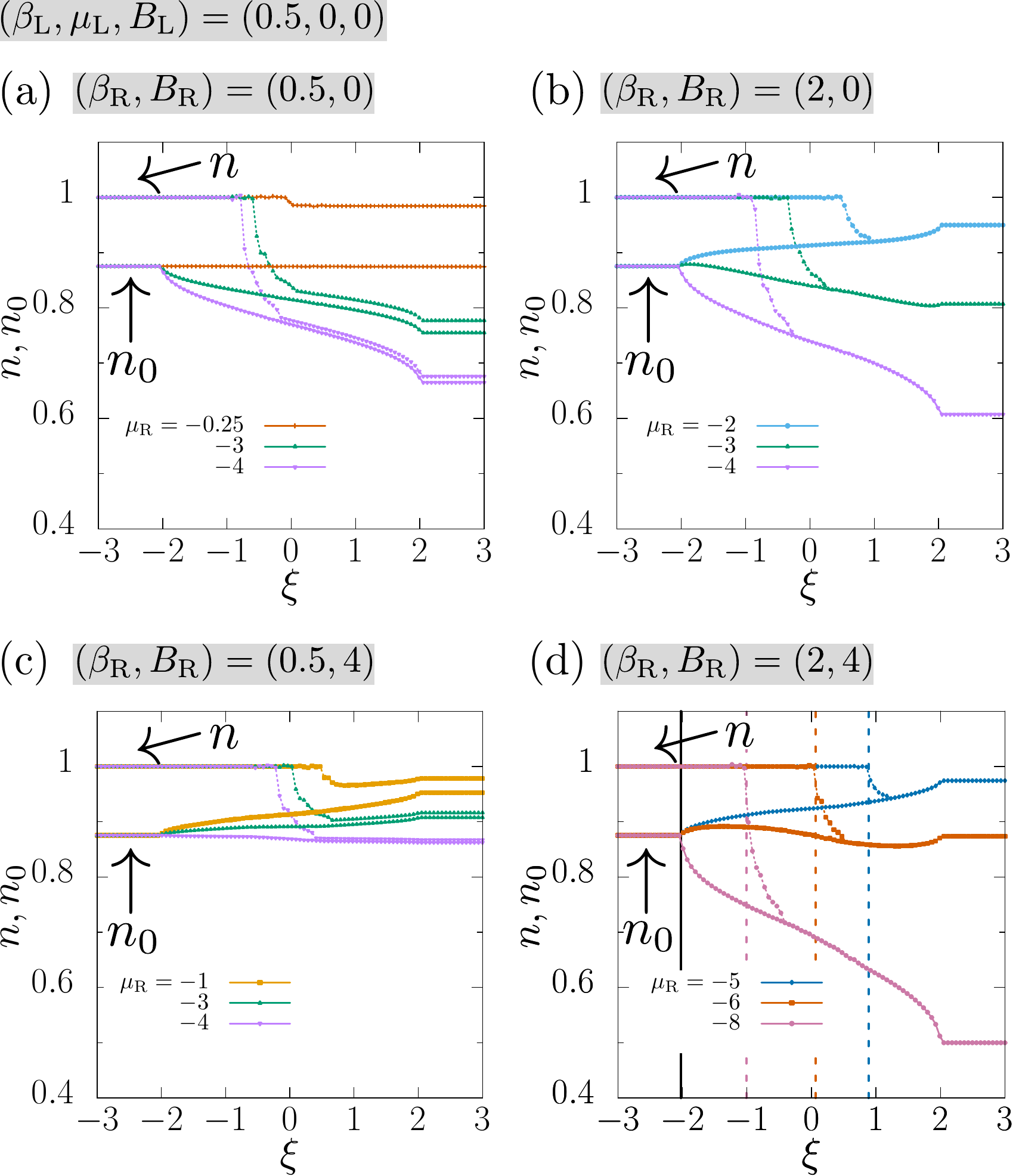}
\caption{
Real-$k$ quasiparticle density $n_{0}(\xi)$ and 
total electron density $n(\xi)$ 
for four different parameter sets of the right initial state [(a), (b), (c), and (d)]. 
Panel (d) also shows $V_{\LL}$ (solid line) and 
$\xi^{-}_{-\infty}$ (dashed lines).
}
\label{fig:nzeroclogged}
\end{figure}

Figures~\ref{fig:njnclogged} and \ref{fig:njncloggedmagnetic} show 
that stationary charge clogging occurs 
when the initial right state is prepared either
at low temperature or in a large magnetic field. 
We examined the main effect of these initial conditions 
and found that one common effect 
is the high density of real-$k$ quasiparticle excitations 
$n^{\RR}_{0}$.  
This is because the real-$k$ excitations have an energy 
lower than charge bound states ($k$-$\clam$ string) and 
have spin $1/2$, 
while $k$-$\clam$ string carries spin $0$. 
Therefore, real-$k$ excitations have a higher density 
at lower temperature, and they are more susceptible to magnetic field.

To study this point systematically, 
we fix the left initial state and vary $\mu_{\RR}$ to check stationary clogging 
for each $(\beta_{\RR}, B_{\RR})$ of four choices. 
Figure~\ref{fig:nzeroclogged} shows the calculated
profiles of $n(\xi)$ and $n_{0}(\xi)$.
The initial left state is the same one as in Figs.~\ref{fig:njnclogged} 
and \ref{fig:njncloggedmagnetic}: 
$(\beta_{\LL},\mu_{\LL},B_{\LL})=(0.5,0,0)$.    
In this case, $n^{\LL}=1$ and $n^{\LL}_{0}\approx 0.875$.
The four sets of $(\beta_{\RR},B_{\RR})$ are also the same 
as those in Figs.~\ref{fig:njnclogged} 
and \ref{fig:njncloggedmagnetic} 
except for the set $(2,4)$. 
The results show that stationary charge clogging 
for all the $(\beta_{\RR},B_{\RR})$ sets except $(0.5,0)$ 
when $|\mu_{\RR}|$ is small.  

\begin{figure}[tb]
\centering
\includegraphics[width=7cm]{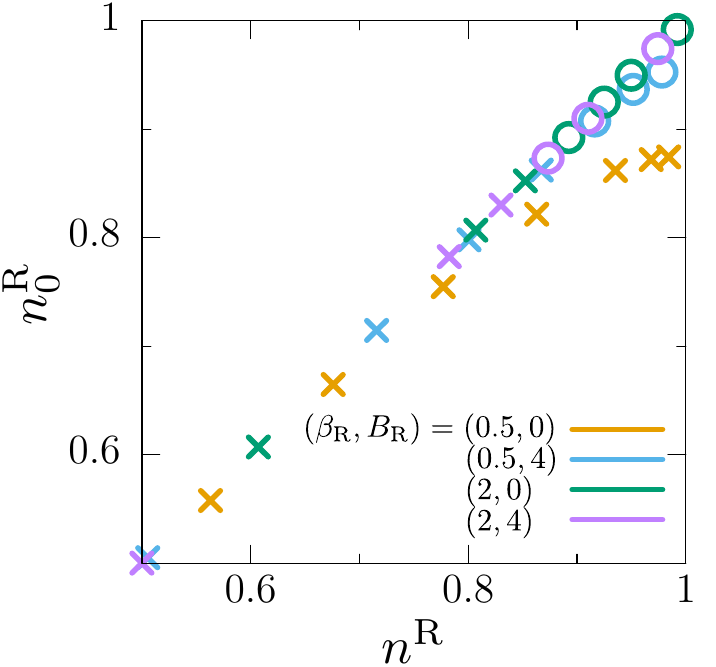}
\caption{
The real-$k$ quasiparticle density $n_{0}^{\RR}$ 
plotted versus the total density $n^{\RR}$ 
in the initial right state.  
$\mu_{\RR}$ is varied 
in the range $-8\leq \mu_{\RR}\leq -0.25$
for each set of $(\beta_{\RR},B_{\RR})$. 
Circles show that stationary charge clogging occurs, 
while crosses show no clogging.
}
\label{fig:nrzvsnrcc}
\end{figure}

Let us first examine how stationary charge clogging 
correlates with the total electron density $n^{\RR}$
and real-$k$ quasiparticle density $n_{0}^{\RR}$
in the initial right state.
Figure~\ref{fig:nrzvsnrcc} shows $n_{0}^{\RR}$ and 
$n^{\RR}$ 
and also shows whether stationary charge clogging occurs. 
The cases that stationary charge clogging occurs are 
shown by a circle symbol.  
The parameters $(\beta_{\RR},B_{\RR})$ for 
the right initial state are identical to those used 
in Fig.~\ref{fig:nzeroclogged}, 
but a larger number of $\mu_{\RR}$ values are used.  
This plot shows that the most important factor 
for realizing stationary charge clogging is a high density 
of $n_{0}^{\RR}$.
As shown by the results for $(\beta_{\RR},B_{\RR})=(0.5,0)$, 
the total density $n^{\RR}$ is large but 
stationary clogging does not occur.  
Therefore, $n^{\RR}$ is not a primary factor 
to determine the appearance of stationary charge clogging.  

\begin{figure}[tb]
\centering
\includegraphics[width=7cm]{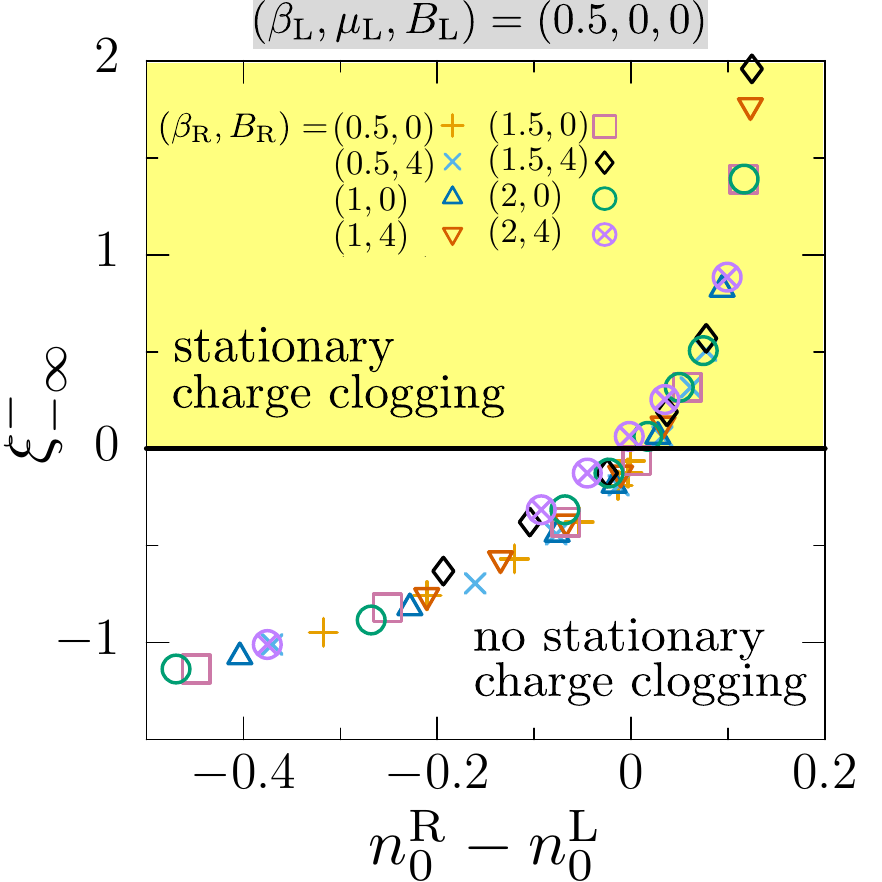}
\caption{
Right border of charge clogged region 
$\xi_{-\infty}^{-}$ plotted versus 
the density difference of real-$k$ quasiparticles 
between the two initial states 
$n_{0}^{\RR}-n_{0}^{\LL}$. 
The part of $\xi_{-\infty}^{-}>0$ is the region of stationary charge clogging.}
\label{fig:clxi}
\end{figure}

We confirm this expectation that stationary charge clogging 
is determined the density of real-$k$ quasiparticles.
Figure~\ref{fig:clxi} shows 
the right border of charge clogged region $\xi_{-\infty}^{-}$ 
plotted versus $n_{0}^{\RR}-n_{0}^{\LL}$. 
The data are calculated  
for the same sets of initial 
conditions as in Fig.~\ref{fig:nrzvsnrcc} 
and supplemented by the results at intermediate temperatures 
$(\beta_{\RR},B_{\RR})=(1,0)$, $(1,4)$, 
$(1.5,0)$, and $(1.5,4)$.  
The part of $\xi_{-\infty}^{-} >0$ corresponds to the cases of stationary 
charge clogging, and this agrees precisely with 
the region of $n_{0}^{\RR}-n_{0}^{\LL} > 0$.  
The results show a universal curve for the different sets 
of data, and this means that the right border $\xi_{-\infty}^{-}$ 
is determined by the one factor $n_{0}^{\RR}-n_{0}^{\LL} > 0$ alone, 
at least when the initial left state is fixed.  
Therefore, although this analysis is based on 
numerical calculations with eight parameter sets,
it is likely that this is a general 
criterion for realizing stationary charge clogging,  
represented explicitly as
\begin{equation}
 ( n^{\LL}-n^{\RR} ) ( n_{0}^{\LL}-n_{0}^{\RR} ) < 0  , 
\mbox{~and~} 
 ( 1 - n^{\LL} ) ( 1 - n^{\RR} ) = 0 , 
\label{eq:cond_clog}
\end{equation}
where the latter condition is equivalent to 
$\mu_{\LL}=0$ or $\mu_{\RR}=0$.
Namely, the majority-minority relation should be reversed 
between the total electron density and the density of 
real-$k$ quasiparticles.  

\begin{figure}[tb]
\centering
\includegraphics[width=\columnwidth,clip]{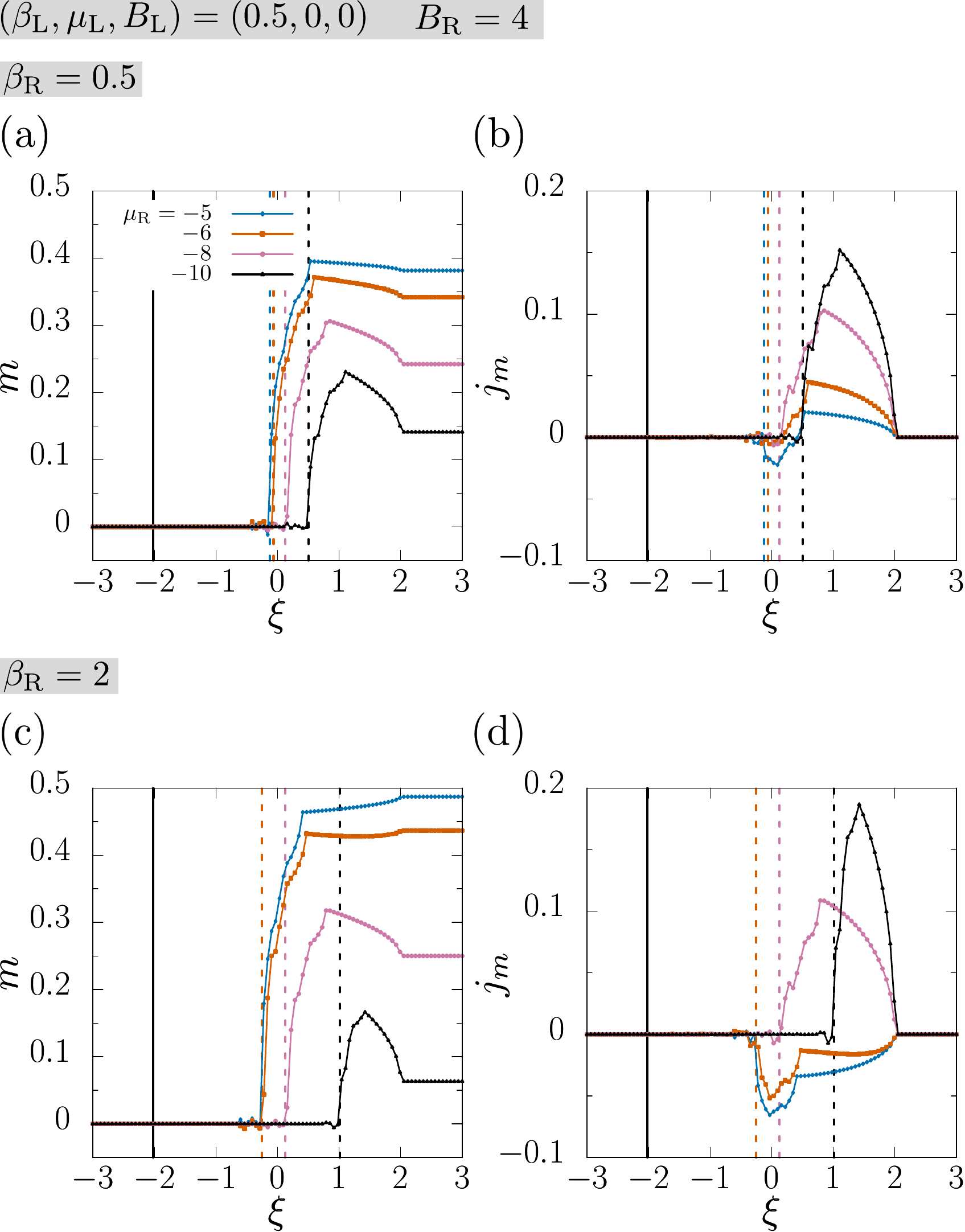}
\caption{
Profiles of 
(a), (c) magnetization density $m(\xi)$ and (b), (d) spin current $j_{m}(\xi)$ 
for various values of chemical potential in the right initial 
state $\mu_{\RR}$.  
Vertical solid and dashed lines represent $V_{\LL}$ and $\xi^{-}_{\infty}$, 
which are the borders of spin clogged region. In (c) and (d), 
$\xi^{-}_{\infty}$'s for $\mu_{\RR}=-5$ and $\mu_{\RR}=-6$ overlap.}
\label{fig:mjmclogged}
\end{figure}

We also examine the possibility of \textit{stationary spin clogging}.  
This occurs when $B_{\LL}=0$ and is characterized as 
\begin{align}
    m (\xi)=0 , \ j_m (\xi)=0 , \ j_{e} (\xi) \not\equiv 0, 
   \quad 
   \bigl( V_{\LL} < {}^\forall \xi < \xi_{\infty}^{-} \bigr),
\end{align}
while $ m \not\equiv 0$, $j_m \not\equiv 0 $ 
at $ \xi_{\infty}^{-} < \xi < V_{\RR}$.  
Figure~\ref{fig:mjmclogged} shows the profiles of 
magnetization $m(\xi)$ and spin current $j_{m}(\xi)$ 
for $-10 \le \mu_{\RR} \le -5$. 
The two sets of initial conditions are those used 
in Figs.~\ref{fig:njncloggedmagnetic} and \ref{fig:nzeroclogged}, 
and $\beta_{\RR}$ is different between the two.  
The right boundary of spin clogged region is $\xi^{-}_{\infty}$ 
not $\xi^{-}_{-\infty}$, and it is shown by dashed lines. 
For both sets of initial conditions, stationary spin clogging 
occurs for $\mu_{\RR} \le -8$. 
In contrast to charge clogging, 
stationary spin clogging occurs 
when $|\mu_{\RR}|$ is large, i.e., $n^{\RR}$ is small. 
Similar to stationary charge clogging, 
we numerically confirmed that stationary spin clogging occurs, 
if the following criterion is satisfied
\begin{equation}
 ( m^{\LL}-m^{\RR} ) ( m_{0}^{\LL}-m_{0}^{\RR} ) < 0 , 
\mbox{~and~} 
 m^{\LL}m^{\RR} = 0 . 
\label{eq:cond_spinclog}
\end{equation}
For all the initial conditions used, we never 
found the coexistence of charge and spin cloggings  
in the stationary state. 

Finally, we note that back current is not limited to the case 
when stationary charge clogging occurs. 
Figure~\ref{fig:backcurrent} shows an example 
in the case of $(\beta_{\LL},\mu_{\LL},B_{\LL})=(0.5,-0.25,0)$ 
and $(\beta_{\RR},\mu_{\RR},B_{\RR})=(2,-5,4)$. 
In this case, 
$n^{\LL}\approx 0.984 > n^{\RR} \approx 0.974 \approx n_0^{\RR} > n_0^{\LL} \approx 0.875$. 
Both initial states are prepared with nonzero chemical potential, 
and stationary charge clogging does not occur. 
However, Fig.~\ref{fig:backcurrent}(b) shows that back current ($j_n < 0$) flows. 
As in the cases of stationary charge clogging, 
$n(\xi)$ is not monotonic while $n_{0}(\xi)$ is monotonic. 
In the region $V_{\LL} < \xi < \xi_{-\infty}^{-}$, 
$n(\xi )$ increases slightly with $\xi$ and approaches 
toward half filling. 
The current $j_n (\xi )$ is nonzero but its amplitude is small. 
In this sense, this is a \textit{``pseudo clogged'' region}.  
For $\xi > \xi_{-\infty}^{-}$, 
$n(\xi )$ decreases quickly and this is accompanied 
by a large enhancement of $j_n$.  
This decrease stops at $\xi = V_{\RR , 1}=\xi_{-1}^{+}$ 
and then $n (\xi)$ increases again.  

It is notable that 
back current extends over the entire transient 
region $V_{\LL} < \xi < V_{\RR}$. 
Therefore, the stationary particle density $n(0)$ 
is larger than the initial particle densities in both left 
and right parts $n(0) > \max \{ n^{\RR}, n^{\LL} \}$ 
as shown in Fig.~\ref{fig:backcurrent}(a). 
One should note that the back current is attributed to 
real-$k$ quasiparticle current $j_{n,0} <0$, 
and all the contributions of bound state excitations 
show a normal behavior, i.e., flows towards lower-density side $j_{n,a}>0$ for $a<0$.

\section{Proportionality of currents}
\label{sec:5} 
In this section, 
we investigate the relations, particularly proportionality, 
among spin, charge, and energy currents.  
As will be shown later, there exist several 
$\xi$-regions showing different behaviors of current ratios.  
To realize nonzero spin current as well as charge current, 
we use initial conditions that nonzero magnetic 
field and chemical potential are applied to the 
left initial state: 
$\bar{\mu}\equiv \beta_{\LL}\mu_{\LL}\neq 0$ and 
$\bar{B}\equiv \beta_{\LL}B_{\LL}\neq 0$.  
In our previous work~\cite{PhysRevB.101.035121}, 
we studied the ratio of spin and charge currents 
in the high-temperature limit $\beta_{\LL}=0$ 
and showed that the following simple relation holds 
in a region connected to the left thermal equilibrium 
\begin{equation}
  \frac{ j_{m}\paren{\xi} }{  j_{n}\paren{\xi} }
 =\frac{ \tanh \bar{B} }{ 2 \tanh |\bar{\mu}| }  , 
\ \ \ 
( \beta_{\LL} \rightarrow 0, \ V_{\LL} < \xi < V_{\LL ,1}),
\label{eq:jnandjm}
\end{equation}
where $V_{\LL , 1}$ is defined in Eq.~\eqref{eq:defboundary}.

\begin{figure}[tb]
\centering 
\includegraphics[width=\columnwidth,clip]{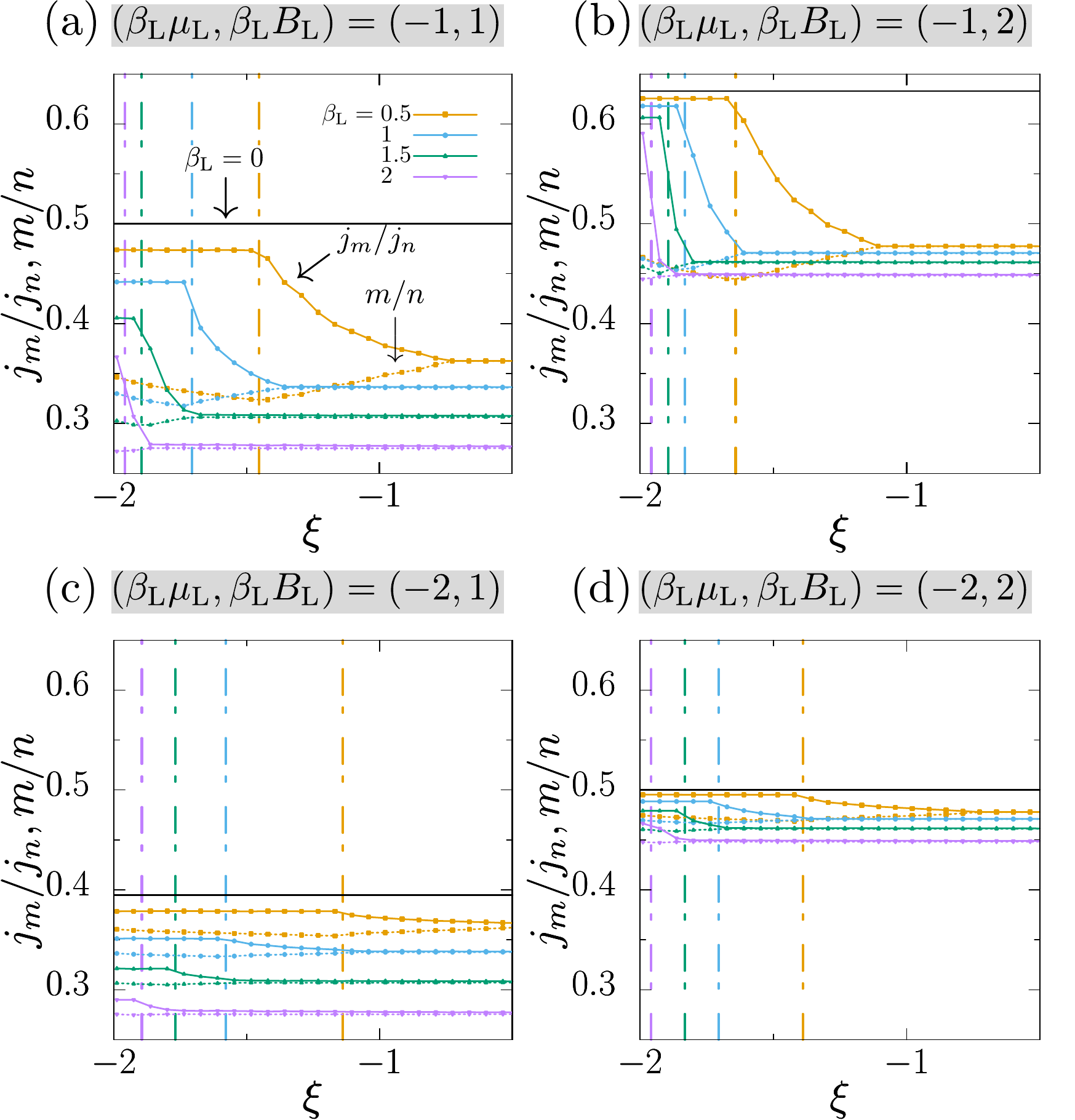}
\caption{
Profiles of the current ratio $j_{m}(\xi)/j_{n}(\xi)$ (solid lines) and 
the corresponding density ratio  $m(\xi)/n(\xi)$~(dashed lines).  
In each panel [(a), (b), (c), and (d)], 
the temperature in the initial left state is varied 
in the range $0.5\leq\beta_{\LL}\leq 2$, while the pair 
$(\beta_{\LL} \mu_{\LL}, \beta_{\LL} B_{\LL})$ is fixed.  
The initial right state is an electron vacuum $n^{\RR}=0$. 
Vertical dashed-dotted line shows $V_{\LL ,1}$, 
while horizontal black solid line represents 
the high-temperature limit (\ref{eq:jnandjm}). }
\label{fig:jmovjntempdep}
\end{figure}

Let us first investigate how nonzero $\beta_{\LL}$ 
changes this proportionality.
Figure~\ref{fig:jmovjntempdep} shows the spatial profiles of 
the current ratio $j_{m}(\xi)/j_{n}(\xi)$ and the corresponding
density ratio $m(\xi)/n(\xi)$ for various values of $\beta_{\LL}$ 
whereas $(\bar{\mu}, \bar{B})$ is fixed to one of four pairs.  
Note that $\mu_{\LL}$ and $B_{\LL}$ are also varied simultaneously 
to fix $(\bar{\mu}, \bar{B})$.  
The initial right state is set so that $n^{\RR}=0$.  
All the data in Figs.~\ref{fig:jmovjntempdep}(a)-\ref{fig:jmovjntempdep}(d) show a \textit{plateau behavior} 
of $j_{m} (\xi )/j_{n} (\xi )$.  
Namely, the region of constant current ratio persists 
for all $\beta_{\LL} > 0$'s used  
and its width agrees quite well with 
$V_{\LL} < \xi < V_{\LL ,1}$ calculated for each $\beta_{\LL}$.  
For $\xi > V_{\LL , 1}$, the current ratio decreases with $\xi$, 
but shows a second plateau for larger $\xi$, 
which will be discussed later in detail.  
Varying $\beta_{\LL}$ at least in the high-temperature region 
does not destroy a plateau in the current ratio but has 
two main effects.  
The first effect is about the constant ratio of $j_{m}/j_{n}$. 
Lowering temperature with keeping $(\bar{\mu}, \bar{B})$ fixed 
decreases its value from 
the high-temperature limit (\ref{eq:jnandjm}), which is 
shown by the black solid line in each panel of the figure.  
The second effect is about the width of the first plateau, 
and lowering temperature shrinks its width.  
Comparison of the data for the different sets of $(\bar{\mu}, \bar{B})$ 
shows that 
larger $|\mu_{\LL}|$ or smaller $|B_{\LL}|$ expands the plateau width 
when $\beta_{\LL}$ is fixed.  

It is important that the plateau value of the current ratio 
$j_{m}/j_{n}$ differs from the corresponding density ratio 
$m (\xi )/n (\xi )$, which changes with $\xi$ in this region,
and also from 
the ratio in the left equilibrium state $ m^{\LL} / n^{\LL} $. 
This is consistent with the fact that these ratios differ 
from Eq.~\eqref{eq:jnandjm} 
in the high-temperature limit~\cite{PhysRevB.101.035121,10.1143/PTP.47.69}: 
\begin{align}
\frac{m (\xi )}{n (\xi )} 
&= \frac{\tanh \bar{B}}{2 \tanh |\bar{\mu}|} 
\left[ 1 - \frac{2 n (\xi )^{-1} }{e^{2 |\bar{\mu}|}+1} 
\right], 
\ \ \ 
(V_{\LL} < \xi < V_{\LL ,1}) ,
\\
\frac{m^{\LL}}{n^{\LL}} 
&= \frac{\sinh \bar{B}}{2 (e^{-|\bar{\mu}|} + \cosh \bar{B})} . 
\end{align}
The current and density ratios, $j_{m}/j_{n}$ and $m/n$, 
become closer with further increasing $\xi > V_{\LL , 1} $.  

\begin{figure}[tb]
\centering 
\includegraphics[width=\columnwidth,clip]{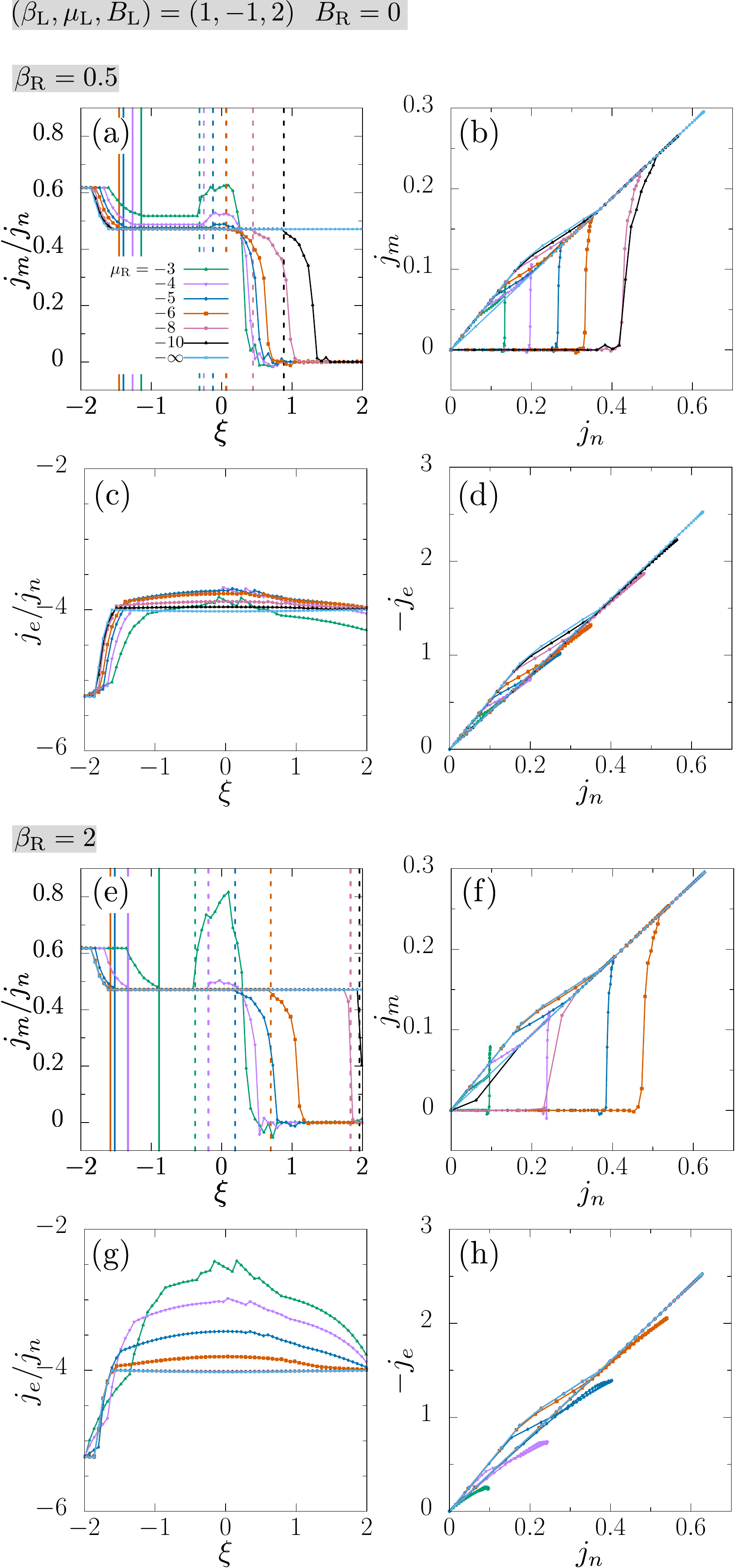}
\caption{
Effects of varying the initial right conditions on the profiles
of the ratios (a), (e) $j_m / j_n$ and (c), (g) $j_e /j_n$. 
Characteristics are also shown for (b), (f) $j_n$-$j_m$ and 
(d), (h) $j_n$-$j_e$.  
In panels (a) and (e), vertical solid lines represent $\xi_{-1}^{+}$, and vertical dashed lines represent $\xi_{1}^{-}$.}
\label{fig:rightdep}
\end{figure}

We next examine how the right initial state changes 
the proportionality among $j_{n}$, $j_{m}$, and energy current $j_{e}$. 
Figure~\ref{fig:rightdep} shows the results for the two values of 
temperature, $\beta_{\RR}=0.5$ and 2.  
For each value, chemical potential is varied in the 
range $ -\infty \le \mu_R \le -3$,
and the upper two panels show the proportionality between $j_{m}$ and $j_{n}$, 
while the lower two are for $j_{e}$ and $j_{n}$. 
The other parameters of the initial condition are set
to the same values used in Fig.~\ref{fig:jmovjntempdep}(b): 
$(\beta_{\LL},\mu_{\LL},B_{\LL})=(1,-1,2)$ and $B_{\RR}=0$. 

Let us discuss the results in Fig.~\ref{fig:rightdep}.
Figures \ref{fig:rightdep}(a) and \ref{fig:rightdep}(e) show that the ratio $j_{m}/j_{n}$
in the first plateau region $V_{\LL} < \xi < V_{\LL , 1}$ hardly depends
on the initial right conditions $j_{m}/j_{n}\approx 0.62$.
This is also reflected by a universal slope of the lines 
starting from the origin in Figs.~\ref{fig:rightdep}(b) and \ref{fig:rightdep}(f). 
In all of these cases, 
$V_{\LL , 1}=\xi_{-1}^{-}$, and 
this means that the right boundary of the current ratio plateau 
is characterized by the left light cone of the charge bound state 
with $a=-1$. 
Thus, quasiparticles with charge $2e$ break 
the constant current ratio. 

Another interesting finding is that the current ratio $j_{m}/j_{n}$ 
shows a second plateau around $\xi \sim -1$ 
in Figs.~\ref{fig:rightdep} (a) and (e). 
The $j_{m}$-$j_{n}$ curves in the panels (b) and (f) show 
this second plateau as an almost straight inclined line 
in the most distant part from the origin.  
The current ratio in the second plateau is smaller than the 
value in the first plateau.  
In contrast to the first plateau, this value depends on the 
initial condition $\mu_{\RR}$ in the right part, and 
this is particularly evident at $\beta_{\RR}=0.5$.  
The ratio increases as $\mu_{\RR}$ approaches zero, and this 
control corresponds to varying $n_{\RR}$ towards $n_{\LL}$.  
We found that the left border of the second plateau 
is determined by the light cone $\xi_{-1}^{+}$. 
Its value is calculated for each $\mu_{\RR}$ 
and shown by a solid line in the figure 
($-6\leq \mu_{\RR}\leq -3$ for visibility). 
The right border is the light cone $\xi_{1}^{-}$ 
and shown by a dashed line.   
It is interesting that the second plateau expands as 
$\mu_{\RR}$ goes down ($n_{\RR}$ decreases). 
This is related to the size of spin clogged region, which 
will be explained below.   

With further increasing $\xi$, the second plateau terminates 
and the current ratio $j_{m}/j_{n}$ shows a continuous 
drop down to 0.  
This part is represented in the $j_{m}$-$j_{n}$ curves as 
a vertical edge of each triangular loop.  

The rightmost part connected to the right initial state 
is a spin clogged region, where $j_{m}/j_{n} =0$.  
This is dual to charge clogged region, which appears when 
the two parts with $\mu =0$ and $\ne 0$ are connected. 
In this case, the two parts with $B=0$ and $\ne 0$ are connected 
and a spin clogged region appears.  
The spin clogged region corresponds to a base of 
each $j_{m}$-$j_{n}$ loop.  
Its width is determined by $V^{\RR}-\xi_{\infty}^{+}$ and expands 
as $\mu_{\RR}$ goes up (i.e., higher density of $n_{\RR}$).  

\begin{figure}[tb]
\centering
\includegraphics[width=7cm]{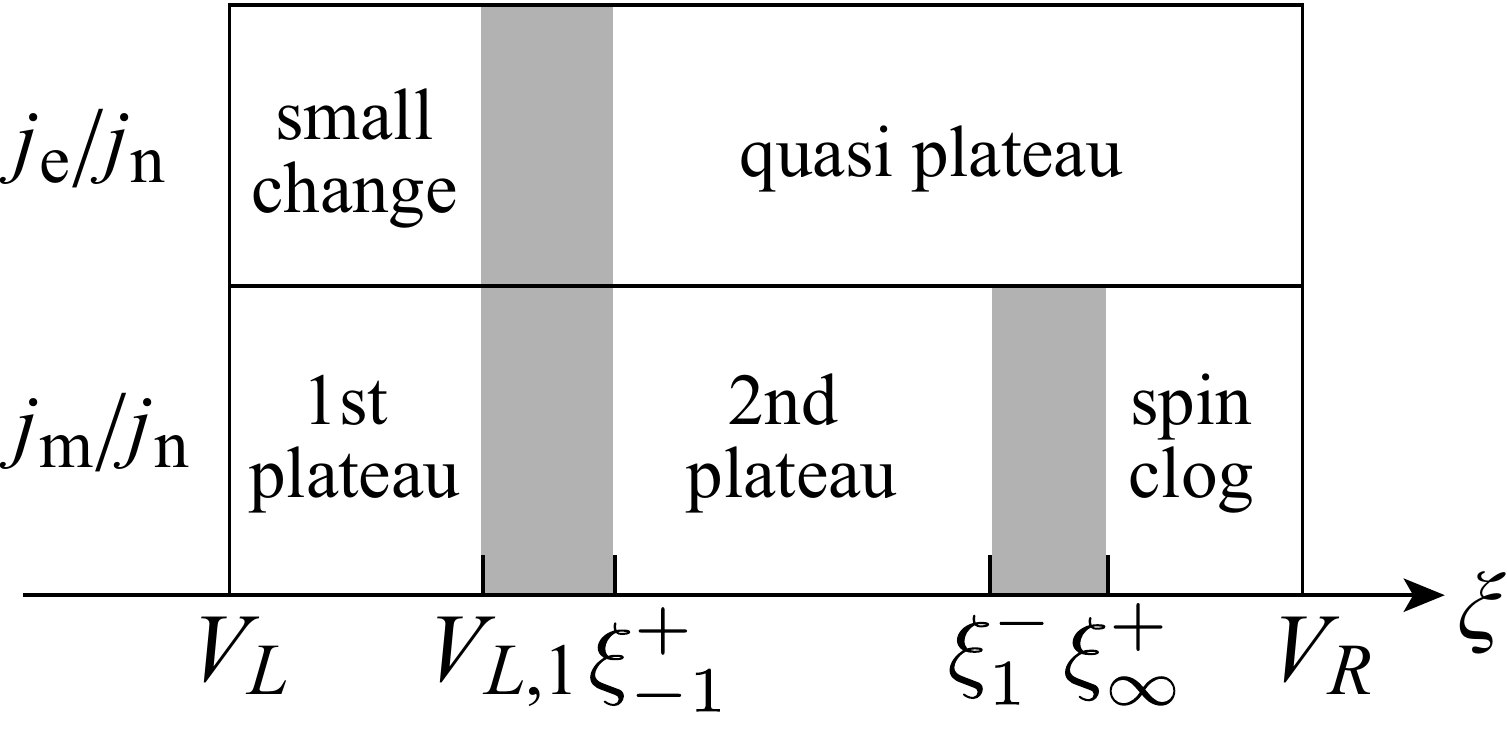}
\caption{
Summary of the behavior of the current ratios 
$j_{m}(\xi ) / j_{n} (\xi )$ and $j_{e}(\xi ) / j_{n} (\xi )$, 
for the initial conditions with 
$\mu_{\LL} \ne 0$, $B_{\LL} \ne 0$, and $B_{\RR} =0$.  
Shaded parts are transient regions where 
the current ratio shows a large change.}
\label{fig:current-prop}
\end{figure}

Thus, summarizing the behavior of the current ratio $j_m / j_n$,  
the whole transient space $V_{\LL} < \xi < V_{\RR}$ is divided 
into five regions as shown in Fig.~\ref{fig:current-prop}: two plateaus and one spin clogged regions  
separated by two transient regions.  
This is different when considering 
the ratio of energy and charge currents $j_{e} / j_{n} $.

Figure \ref{fig:rightdep} also shows the proportionality 
between energy and particle density currents. 
Figures \ref{fig:rightdep}(c) and \ref{fig:rightdep}(g) show that the ratio $j_{e}/j_{n}$ 
changes with $\xi$ and its value also varies 
with the initial right conditions in the $\xi$-region 
of the first plateau of $j_{m} / j_{n}$.  
However, as $\xi$ approaches $V_{\LL}$, 
the ratio $j_{e}/j_{n}$ approaches
a universal value $\sim -5.2$.  
This value is independent of the initial right conditions.  
We calculated the ratio for other initial conditions 
and found that it depends on the initial left conditions. 
This result means that near the left thermal equilibrium state
each carrier in particle density current
also carries the identical energy irrespective 
of the initial right conditions.

For $\xi > V_{\LL , 1}$, 
the ratio $j_{e}/j_{n}$ shows a large $\xi$ dependence for a while,  
but Figure~\ref{fig:rightdep}(c) shows that 
$\xi$-dependence is strongly suppressed 
in the region of the second plateau of $j_{m}/j_{n}$ and 
also at other $\xi < V_{\RR}$.  
This quasi-plateau behavior becomes more evident as $\mu_{\RR}$ 
decreases, and the $j_{e}$-$j_{n}$ curve shows a retracing 
straight line in Fig.~\ref{fig:rightdep}(d).  
The corresponding results for the case of lower temperature 
in the initial right state $\beta_{\RR}=2$ are plotted 
in Figs.~\ref{fig:rightdep}(g) and \ref{fig:rightdep}(h). 
In this case, the ratio $j_{e}/j_{n}$ shows a large 
$\xi$ dependence in the whole $\xi$ space, particularly 
for $\mu_{\RR}=-3$ and $-4$, but the amplitude of both currents 
is quite small in those cases as shown in Fig.~\ref{fig:rightdep}(h).  
Decreasing $\mu_{\RR}$ below $-4$, the $\xi$-dependence 
in the region $\xi > -1.5$ is suppressed as in the case of 
$\beta_{\RR}=0.5$, and the $j_{e}$-$j_{n}$ curve shows a 
quite straight path in Fig.~\ref{fig:rightdep}(h).  
With decreasing $\mu_{\RR}$, the ratio $j_{e}/j_{n}$ 
converges $-4$ in the quasi-plateau region. 
This value corresponds to the energy of a carrier 
with the fastest velocity 
$\max_{k} \ddd{v}{0}^{\RR}(k)=\ddd{v}{0}^{\RR}(\pi/2)=2$.
As shown in Eq.~\eqref{eq:be}, that is $e_0(\pi/2)=-2u=-4$, 
since $u=2$ in the present work.

Thus, summarizing the behavior of the ratio $j_e / j_n$,  
the whole transient space $V_{\LL} < \xi < V_{\RR}$ is 
now divided into three regions as shown in Fig.~\ref{fig:current-prop}.  
In the first plateau and the first transient regions of $j_{m}/j_{n}$, 
the ratio $j_{e} (\xi )/j_{n} (\xi )$ shows a small and large 
dependence on $\xi$, respectively.  
In the remaining region, 
the ratio $j_{e} (\xi )/j_{n} (\xi )$ shows a quasi-plateau 
behavior that becomes evident as $\mu_{\RR}$ decreases.  

\section{Conclusions}
\label{sec:6}

In this paper, we mainly studied two issues of
the nonequilibrium quench dynamics in the 1D Hubbard model
based on the generalized hydrodynamics
theory with the partitioning protocol.

The first issue is the possibility of charge and spin clogging
in a stationary state, i.e., the phenomenon where 
charge or spin current is zero whereas nonvanishing
energy current flows at the ray $\xi = x/t =0$. 
We examined various cases of initial conditions 
under the constraint that the initial particle density 
is $n^{\LL}=1$ and $n^{\RR}<1$ for the two parts. 
  
When stationary charge clogging occurs,
the half-filled region expands to the right side.
In Sec.~\ref{sec:4}, we numerically solved the GHD equations
for various initial conditions and found several cases
of stationary charge and spin clogging.  
We studied the dependence on the initial conditions and
found that an important factor is the particle density
of scattering states $n_{0}$.
We found by numerical calculations that the condition $n_{0}^{\LL}<n_{0}^{\RR}$ 
is crucial for stationary charge clogging 
[more precise one in Eq.~\eqref{eq:cond_clog}]. 
Similar to stationary charge clogging, when stationary spin clogging occurs,  
$m_{0}^{\LL}-m_{0}^{\RR}>0$ is satisfied in the cases of $m^{\LL}=0$ and $m^{\RR}>0$. 
When stationary charge clogging occurs, $n(\xi)$ 
should be nonmonotonic, 
and we found that there exists a back current, 
which flows towards the higher-density region ($n=1$). 
In the right part, as time goes, 
the particle density decreases first and then increases 
to be half filling, while the sign of its current does not change $j_{n}<0$.

The second issue is the proportionality among currents of 
spin $j_m$, charge $j_n$, and energy $j_e$. 
The current ratio $j_m /j_n$ in the high-temperature limit 
$\beta_{\LL}=0$
was studied in our previous work~\cite{PhysRevB.101.035121}.  
We numerically studied this issue
at finite temperatures in Sec.~\ref{sec:5} 
and calculated the profiles of 
spin, particle density (equivalent to charge), and energy currents
for various sets of initial conditions and analyzed the results.  
We found that the constant proportionality of $j_m$ and $j_n$ 
in the region $V_{\LL}<\xi<V_{\LL,1}$ persists 
even at finite temperatures (the first plateau).  
The value of this constant ratio depends on the initial left 
temperature $\beta_{\LL}$, but its dependence 
on the initial right conditions is negligible. 
Another finding is a second plateau of $j_m / j_n$ 
in the region of $\xi_{-1}^{+}<\xi<\xi_{1}^{-}$. 
In contrast to the first plateau, the constant ratio 
in the second plateau depends on the initial right conditions.

We also analyzed the ratio of energy and charge current 
$j_e / j_n$ with controlling the initial right conditions. 
We found that as $\xi$ approaches the left end of 
the transient region $ V_{\LL}$  
the ratio, $j_e / j_n$ approaches a constant value 
independent of the initial right conditions.  
In a wide $\xi$-region including the second plateau of $j_m /j_n$, 
the ratio $j_e / j_n$ also shows a quasi-plateau behavior 
particularly when $\mu_{\RR}$ is not so small. 
When $|\mu_{\RR}|$ is large, the ratio $j_e / j_n$ 
in the quasi plateau approaches the universal value $-2u$. 

\section*{Acknowledgments}
Calculations in this work were partly performed using
the facilities of the Supercomputer Center at ISSP,
the University of Tokyo.


\begin{thebibliography}{99}

\bibitem{Calabrese_2016}
P.~Calabrese, F.~H.~L. Essler, and G.~Mussardo,
\newblock Introduction to `Quantum Integrability in Out of Equilibrium
  Systems',
\newblock J. Stat. Mech. {\bf 2016}, 064001 (2016).

\bibitem{PhysRevX.6.041065}
O.~A. Castro-Alvaredo, B.~Doyon, and T.~Yoshimura,
\newblock Emergent Hydrodynamics in Integrable Quantum Systems Out of
  Equilibrium,
\newblock Phys. Rev. X {\bf 6}, 041065 (2016).

\bibitem{PhysRevLett.117.207201}
B.~Bertini, M.~Collura, J.~De~Nardis, and M.~Fagotti,
\newblock Transport in Out-of-Equilibrium $XXZ$ Chains: Exact Profiles of
  Charges and Currents,
\newblock Phys. Rev. Lett. {\bf 117}, 207201 (2016).

\bibitem{PhysRevLett.122.090601}
M.~Schemmer, I.~Bouchoule, B.~Doyon, and J.~Dubail,
\newblock Generalized Hydrodynamics on an Atom Chip,
\newblock Phys. Rev. Lett. {\bf 122}, 090601 (2019).

\bibitem{rubin1971abnormal}
R.~J. Rubin and W.~L. Greer,
\newblock Abnormal Lattice Thermal Conductivity of a One-dimensional, Harmonic,
  Isotopically Disordered Crystal,
\newblock J. Math. Phys. {\bf 12}, 1686 (1971).

\bibitem{spohn1977stationary}
H.~Spohn and J.~L. Lebowitz,
\newblock Stationary non-equilibrium states of infinite harmonic systems,
\newblock Commun. Math. Phys. {\bf 54}, 97 (1977).

\bibitem{bernard2012energy}
D.~Bernard and B.~Doyon,
\newblock Energy flow in non-equilibrium conformal field theory,
\newblock J. Phys. A {\bf 45}, 362001
  (2012).

\bibitem{Bernard2015}
D.~Bernard and B.~Doyon,
\newblock Non-equilibrium steady states in conformal field theory,
\newblock Ann. Henri Poincar{\'e} {\bf 16}, 113 (2015).

\bibitem{bhaseen2015energy}
M.~J. Bhaseen, B.~Doyon, A.~Lucas, and K.~Schalm,
\newblock Energy flow in quantum critical systems far from equilibrium,
\newblock Nat. Phys. {\bf 11}, 509 (2015).

\bibitem{fagotti2016charges}
M.~Fagotti,
\newblock Charges and currents in quantum spin chains: late-time dynamics and
  spontaneous currents,
\newblock J. Phys. A {\bf 50}, 034005
  (2016).

\bibitem{PhysRevB.96.020403}
A.~De~Luca, M.~Collura, and J.~De~Nardis,
\newblock Nonequilibrium spin transport in integrable spin chains: Persistent
  currents and emergence of magnetic domains,
\newblock Phys. Rev. B {\bf 96}, 020403(R) (2017).

\bibitem{doyon2017dynamics}
B.~Doyon and H.~Spohn,
\newblock Dynamics of hard rods with initial domain wall state,
\newblock J. Stat. Mech. , 073210 (2017).

\bibitem{PhysRevB.97.045407}
V.~B. Bulchandani, R.~Vasseur, C.~Karrasch, and J.~E. Moore,
\newblock Bethe-Boltzmann hydrodynamics and spin transport in the XXZ chain,
\newblock Phys. Rev. B {\bf 97}, 045407 (2018).

\bibitem{PhysRevLett.120.045301}
B.~Doyon, T.~Yoshimura, and J.-S. Caux,
\newblock Soliton Gases and Generalized Hydrodynamics,
\newblock Phys. Rev. Lett. {\bf 120}, 045301 (2018).

\bibitem{PhysRevB.97.081111}
M.~Collura, A.~De~Luca, and J.~Viti,
\newblock Analytic solution of the domain-wall nonequilibrium stationary state,
\newblock Phys. Rev. B {\bf 97}, 081111(R) (2018).

\bibitem{Bertini_2018}
B.~Bertini and L.~Piroli,
\newblock Low-temperature transport in out-of-equilibrium {XXZ} chains,
\newblock J. Stat. Mech. , 033104 (2018).

\bibitem{PhysRevLett.120.176801}
B.~Bertini, L.~Piroli, and P.~Calabrese,
\newblock Universal Broadening of the Light Cone in Low-Temperature Transport,
\newblock Phys. Rev. Lett. {\bf 120}, 176801 (2018).

\bibitem{10.21468/SciPostPhys.4.6.045}
A.~Bastianello, B.~Doyon, G.~Watts, and T.~Yoshimura,
\newblock {Generalized hydrodynamics of classical integrable field theory: the
  sinh-Gordon model},
\newblock SciPost Phys. {\bf 4}, 45 (2018).

\bibitem{PhysRevB.98.075421}
L.~Mazza, J.~Viti, M.~Carrega, D.~Rossini, and A.~De~Luca,
\newblock Energy transport in an integrable parafermionic chain via generalized
  hydrodynamics,
\newblock Phys. Rev. B {\bf 98}, 075421 (2018).

\bibitem{PhysRevB.99.014305}
M.~Mesty\'an, B.~Bertini, L.~Piroli, and P.~Calabrese,
\newblock Spin-charge separation effects in the low-temperature transport of
  one-dimensional Fermi gases,
\newblock Phys. Rev. B {\bf 99}, 014305 (2019).

\bibitem{PhysRevB.99.174203}
U.~Agrawal, S.~Gopalakrishnan, and R.~Vasseur,
\newblock Generalized hydrodynamics, quasiparticle diffusion, and anomalous
  local relaxation in random integrable spin chains,
\newblock Phys. Rev. B {\bf 99}, 174203 (2019).

\bibitem{doi:10.1063/1.5096892}
B.~Doyon,
\newblock Generalized hydrodynamics of the classical Toda system,
\newblock J. Math. Phys. {\bf 60}, 073302 (2019).

\bibitem{Bulchandani_2019}
V.~B Bulchandani, X.~Cao, and J.~E Moore,
\newblock Kinetic theory of quantum and classical Toda lattices,
\newblock J. Phys. A {\bf 52}, 33LT01
  (2019).

\bibitem{PhysRevLett.119.020602}
E.~Ilievski and J.~De~Nardis,
\newblock Microscopic Origin of Ideal Conductivity in Integrable Quantum
  Models,
\newblock Phys. Rev. Lett. {\bf 119}, 020602 (2017).

\bibitem{PhysRevB.96.081118}
E.~Ilievski and J.~De~Nardis,
\newblock Ballistic transport in the one-dimensional hubbard model: The
  hydrodynamic approach,
\newblock Phys. Rev. B {\bf 96}, 081118(R) (2017).

\bibitem{SciPostPhys.3.6.039}
B.~Doyon and H.~Spohn,
\newblock {Drude Weight for the Lieb-Liniger Bose Gas},
\newblock SciPost Phys. {\bf 3}, 039 (2017).

\bibitem{PhysRevB.97.245135}
V.~Alba,
\newblock Entanglement and quantum transport in integrable systems,
\newblock Phys. Rev. B {\bf 97}, 245135 (2018).

\bibitem{Bertini_ent}
B.~Bertini, M.~Fagotti, L.~Piroli, and P.~Calabrese,
\newblock Entanglement evolution and generalised hydrodynamics: noninteracting
  systems,
\newblock J. Phys. A {\bf 51}, 39LT01
  (2018).

\bibitem{PhysRevB.99.045150}
V.~Alba,
\newblock Towards a generalized hydrodynamics description of R\'enyi entropies
  in integrable systems,
\newblock Phys. Rev. B {\bf 99}, 045150 (2019).

\bibitem{10.21468/SciPostPhys.7.1.005}
V.~Alba, B.~Bertini, and M.~Fagotti,
\newblock {Entanglement evolution and generalised hydrodynamics: interacting
  integrable systems},
\newblock SciPost Phys. {\bf 7}, 005 (2019).

\bibitem{PhysRevB.96.115124}
L.~Piroli, J.~De~Nardis, M.~Collura, B.~Bertini, and M.~Fagotti,
\newblock Transport in out-of-equilibrium XXZ chains: Nonballistic behavior and
  correlation functions,
\newblock Phys. Rev. B {\bf 96}, 115124 (2017).

\bibitem{10.21468/SciPostPhys.5.5.054}
B.~Doyon,
\newblock {Exact large-scale correlations in integrable systems out of
  equilibrium},
\newblock SciPost Phys. {\bf 5}, 054 (2018).

\bibitem{PhysRevLett.121.230602}
E.~Ilievski, J.~De~Nardis, M.~Medenjak, and T.~Prosen,
\newblock Superdiffusion in One-Dimensional Quantum Lattice Models,
\newblock Phys. Rev. Lett. {\bf 121}, 230602 (2018).

\bibitem{PhysRevLett.121.160603}
J.~De~Nardis, D.~Bernard, and B.~Doyon,
\newblock Hydrodynamic Diffusion in Integrable Systems,
\newblock Phys. Rev. Lett. {\bf 121}, 160603 (2018).

\bibitem{PhysRevB.98.220303}
S.~Gopalakrishnan, D.~A.~Huse, V.~Khemani, and R.~Vasseur,
\newblock Hydrodynamics of operator spreading and quasiparticle diffusion in
  interacting integrable systems,
\newblock Phys. Rev. B {\bf 98}, 220303(R) (2018).

\bibitem{10.21468/SciPostPhys.6.4.049}
J.~De~Nardis, D.~Bernard, and B.~Doyon,
\newblock {Diffusion in generalized hydrodynamics and quasiparticle
  scattering},
\newblock SciPost Phys. {\bf 6}, 49 (2019).

\bibitem{PhysRevLett.122.127202}
S.~Gopalakrishnan and R.~Vasseur,
\newblock Kinetic Theory of Spin Diffusion and Superdiffusion in $XXZ$ Spin
  Chains,
\newblock Phys. Rev. Lett. {\bf 122}, 127202 (2019).

\bibitem{Gopalakrishnan16250}
S.~Gopalakrishnan, R.~Vasseur, and B.~Ware,
\newblock Anomalous relaxation and the high-temperature structure factor of XXZ
  spin chains,
\newblock Proc. Natl. Acad. Sci. USA {\bf 116}, 16250
  (2019).

\bibitem{PhysRevB.102.115121}
M.~Fava, B.~Ware, S.~Gopalakrishnan, R.~Vasseur, and S.~A.~Parameswaran,
\newblock Spin crossovers and superdiffusion in the one-dimensional Hubbard
  model,
\newblock Phys. Rev. B {\bf 102}, 115121 (2020).

\bibitem{PhysRevLett.19.1312}
C.~N. Yang,
\newblock {Some Exact Results for the Many-Body Problem in one Dimension with
  Repulsive Delta-Function Interaction},
\newblock Phys. Rev. Lett. {\bf 19}, 1312 (1967).

\bibitem{GAUDIN196755}
M.~Gaudin,
\newblock Un systeme a une dimension de fermions en interaction,
\newblock Phys. Lett. A {\bf 24}, 55  (1967).

\bibitem{PhysRevLett.21.192.2}
E.~H. Lieb and F.~Y. Wu,
\newblock Absence of Mott Transition in an Exact Solution of the Short-Range,
  One-Band Model in One Dimension,
\newblock Phys. Rev. Lett. {\bf 21}, 192 (1968).

\bibitem{essler2005one}
F.~H.~L. Essler, H.~Frahm, F.~G{\"o}hmann, A.~Kl{\"u}mper, and V.~E. Korepin,
\newblock {\textit{The One-Dimensional Hubbard Model}} (Cambridge University Press, Cambridge, 
England, 2005).

\bibitem{PhysRevB.81.020405}
N.~Hlubek, P.~Ribeiro, R.~Saint-Martin, A.~Revcolevschi, G.~Roth, G.~Behr, B.~B\"uchner, and C.~Hess,
\newblock Ballistic heat transport of quantum spin excitations as seen in ${\text{SrCuO}}_{2}$,
\newblock Phys. Rev. B {\bf 81}, 020405(R) (2010).


\bibitem{PhysRevB.58.1261}
A.~Schwartz, M.~Dressel, G.~Gr\"uner, V.~Vescoli, L.~Degiorgi, and T.~Giamarchi,
\newblock On-chain electrodynamics of metallic $(\mathrm{TMTSF}{)}_{2}X$ salts: Observation of Tomonaga-Luttinger liquid response,
\newblock Phys. Rev. B {\bf 58}, 1261 (1998). 

\bibitem{Nature.397.598}
M.~Bockrath, D.~H.~Cobden, J.~Lu, A.~G.~Rinzler, R.~E.~Smalley, L.~Balents, and P.~L.~McEuen, 
\newblock Luttinger-liquid behaviour in carbon nanotubes,
\newblock Nature (London) {\bf 397}, 598 (1999).

\bibitem{Boll1257}
M.~Boll, T.~A.~Hilker, G.~Salomon, A.~Omran, J.~Nespolo, L.~Pollet, I.~Bloch, and C.~Gross,
\newblock Spin- and density-resolved microscopy of antiferromagnetic correlations in Fermi-Hubbard chains,
\newblock Science {\bf 353}, 1257 (2016).

\bibitem{PhysRevB.101.035121}
Y.~Nozawa and H.~Tsunetsugu,
\newblock Generalized hydrodynamic approach to charge and energy currents in
  the one-dimensional Hubbard model,
\newblock Phys. Rev. B {\bf 101}, 035121 (2020).

\bibitem{WF}
See, for example, W. Jones and N. H. March, \textit{Theoretical Solid State Physics}
  Vol. 2 (John Wiley and Sons, New York, 1972).


\bibitem{PhysRevLett.125.070602}
B.~Pozsgay, 
\newblock Algebraic Construction of Current Operators in Integrable Spin Chains, 
\newblock Phys. Rev. Lett. {\bf 125}, 070602 (2020).

\bibitem{PhysRevLett.113.187203}
L.~Bonnes, F.~H.~L. Essler, and A.~M. L\"auchli,
\newblock {``Light-Cone'' Dynamics After Quantum Quenches in Spin Chains},
\newblock Phys. Rev. Lett. {\bf 113}, 187203 (2014).


\bibitem{10.1143/PTP.47.69}
M.~Takahashi,
\newblock {One-Dimensional Hubbard Model at Finite Temperature},
\newblock Prog. Theor. Phys. {\bf 47}, 69 (1972).

\bibitem{PhysRevB.65.165104}
M.~Takahashi and M.~Shiroishi,
\newblock Thermodynamic Bethe ansatz equations of one-dimensional Hubbard model
  and high-temperature expansion,
\newblock Phys. Rev. B {\bf 65}, 165104 (2002).

\end{thebibliography}

\end{document}